%% file: tvlsi_2016.tex
\documentclass[journal]{IEEEtran}
\IEEEoverridecommandlockouts
\usepackage{graphicx}
\DeclareGraphicsExtensions{.pdf,.jpeg,.png}
\usepackage[cmex10]{amsmath}
\usepackage{url}
\usepackage{todonotes}
\usepackage{xcolor}
\usepackage{color}
\usepackage{ifthen}
\usepackage{textcomp}
\usepackage{comment}
\usepackage{xspace}
\usepackage{tabularx}
\usepackage{booktabs}
\usepackage{listings}
\usepackage{paralist}
\usepackage{multirow}
\usepackage[para]{threeparttable}
\usepackage{multirow}
\usepackage{multicol}
\usepackage{todonotes}

\newboolean{arxiv}
\setboolean{arxiv}{true}

\ifthenelse{\boolean{arxiv}} {
\usepackage[all=normal,floats,paragraphs,%
tracking,wordspacing]{savetrees}
}{
\usepackage[subtle]{savetrees}
}

\usepackage[noadjust]{cite} 

\ifthenelse{\boolean{arxiv}} {
\newcommand{\figpath}[1]{./}
\newcommand{\plotpath}[1]{./}
\newcommand{\contpath}[1]{./}
\newcommand{\tabpath}[1]{./}
}{
\newcommand{\figpath}[1]{figures/}
\newcommand{\plotpath}[1]{matlab/plots/}
\newcommand{\contpath}[1]{content/}
\newcommand{\tabpath}[1]{tables/}
}

\usepackage{capt-of}

\newcommand{\figref}[1]{Figure~\ref{#1}}
\newcommand{\secref}[1]{Section~\ref{#1}}
\newcommand{\tabref}[1]{Table~\ref{#1}}

\newcommand{\riscv}{\mbox{RISC-V\,}}

\usepackage{tikz}
\usetikzlibrary{backgrounds}
\usepackage{pgfplots}

\pgfplotscreateplotcyclelist{mylist}{ 
fill=gray!20,draw=black!90,line width=.2pt\\%
fill=blue!60,draw=black!90,line width=.2pt\\%
fill=red!60,draw=black!90,line width=.2pt\\%
fill=orange!60,draw=black!90,line width=.2pt\\%
fill=yellow!60,draw=black!90,line width=.2pt\\%
fill=green!60,draw=black!90,line width=.2pt\\%
fill=white\\%
}

\begin{document}
\bstctlcite{IEEEexample:BSTcontrol}

\title{A near-threshold \riscv core with DSP extensions for scalable IoT Endpoint Devices}

\author{Michael~Gautschi,~\IEEEmembership{Student Member,~IEEE,}
        Pasquale~Davide~Schiavone,~\IEEEmembership{Student Member,~IEEE,}
        Andreas~Traber,
        Igor~Loi,~\IEEEmembership{Member,~IEEE,}
        Antonio~Pullini,~\IEEEmembership{Student Member,~IEEE,}
        Davide~Rossi,~\IEEEmembership{Member,~IEEE,}
        Eric~Flamand,
        Frank~K.~G\"urkaynak,
        and~Luca~Benini,~\IEEEmembership{Fellow,~IEEE}
\thanks{M. Gautschi, D. Schiavone, A. Traber, A. Pullini, E. Flamand, F.K. G\"urkaynak and L. Benini are with the Integrated Systems Laboratory, ETH Z\"urich, Switzerland, e-mail: ({gautschi, schiavone, atraber, pullinia, eflamand, kgf, lbenini}@iis.ee.ethz.ch).}
\thanks{A. Traber is now with Advanced Signal Pursuit (ACP), Z\"urich, Switzerland}
\thanks{D. Rossi, I. Loi, and L. Benini are also with the University of Bologna, Italy}
\thanks{E. Flamand, D. Rossi, and I. Loi are also with GreenWaves Technologies.}
\thanks{Manuscript submitted to TVLSI on August 15, 2016;}}
\ifthenelse{\boolean{arxiv}} {
\markboth{arXiv Preprint}%
}{
\markboth{Journal of TVLSI,~Vol.~XX, No.~YY, Month~2016}%
}
{Shell \MakeLowercase{\textit{et al.}}: Bare Demo of IEEEtran.cls for Journals}

\maketitle

\begin{abstract}
\input{\contpath/00_abstract}
\end{abstract}


\IEEEpeerreviewmaketitle

\section{Introduction}
\label{sec:intro}
\input{\contpath/01_intro}

\section{Related Work}
\label{sec:relwork}
\input{\contpath/02_relwork}

\section{Parallel Ultra-low Power Platform}
\label{sec:pulp}
\input{\contpath/03_pulp}

\section{RISC-V Core Micro-architecture}
\label{sec:core}
\input{\contpath/04_core_archi}

\section{Toolchain Support}
\label{sec:sw}
\input{\contpath/05_sw}

\section{Experimental Results}
\label{sec:results}
\input{\contpath/06_results}

\section{Conclusion}
\label{sec:conclusion}
\input{\contpath/07_conc}

\section*{Acknowledgment}
The authors would like to thank Germain Haugou for the tool and software support. This work was partially supported by the FP7 ERC Advance project
MULTITHERMAN (No: 291125) and Micropower Deep Learning (No: 162524) project funded by the Swiss NSF. 

\bibliographystyle{IEEEtran}
\bibliography{tvlsi_2016,Settings}

\begin{IEEEbiographynophoto}{Michael Gautschi}
received the M.Sc. degree in electrical engineering and information technology from ETH Z\"urich, Switzerland, in 2012. Since then he has been with the Integrated Systems Laboratory, ETH Z\"urich, pursuing a Ph.D. degree. His current research interests include energy-efficient systems, multi-core SoC design, mobile communication, and low-power integrated circuits.
\end{IEEEbiographynophoto}

\begin{IEEEbiographynophoto}{Pasquale Davide Schiavone}
received his B.Sc. (2013) and M.Sc. (2016) in computer engineering from Polytechnic of Turin. Since 2016 he has started his Ph.D. studies at the Integrated Systems Laboratory, ETH Z\"urich. His research interests include datapath blocks design, low-power microprocessors in multi-core systems and deep-learning architectures for energy-efficient systems.
\end{IEEEbiographynophoto}

\begin{IEEEbiographynophoto}{Andreas Traber}
received the M.Sc. degree in electrical engineering and information technology from ETH Z\"urich, Switzerland, in 2015.
He has been working as a research assistant at the Integrated System Laboratory, ETH Z\"urich, and is now working for Advanced Circuit Pursuit AG, Z\"urich, Switzerland. His research interests include energy-efficient systems, multi-core architecture, and embedded CPU design.
\end{IEEEbiographynophoto}

\begin{IEEEbiographynophoto}{Igor Loi}
received the B.Sc. degree in electrical engineering from the University of Cagliari, Italy, in 2005, and the Ph.D. degree from the Department of Electronics and Computer Science, University of Bologna, Italy, in 2010. He currently holds a researcher position in electronics engineering with the University of Bologna.
He current research interests include ultra-low power multicore systems, memory system hierarchies, and ultra low-power on-chip interconnects.
\end{IEEEbiographynophoto}

\begin{IEEEbiographynophoto}{Antonio Pullini}
received the M.S. degree in electrical engineering from Bologna University, Italy. He was a Senior Engineer at iNoCs S.\`a.r.l., Lausanne, Switzerland. His research interests include low-power digital design and networks on chip.
\end{IEEEbiographynophoto}

\begin{IEEEbiographynophoto}{Davide Rossi}
received the PhD from the University of Bologna, Italy, in 2012. He has been a post doc researcher in the Department of Electrical, Electronic and Information Engineering “Guglielmo Marconi” at the University of Bologna since 2015, where he currently holds an assistant professor position. His research interests focus on energy efficient digital architectures in the domain of heterogeneous and reconfigurable multi and many-core systems on a chip.
In this fields he has published more than 30 papers in international peer-reviewed conferences and journals. 
\end{IEEEbiographynophoto}

\begin{IEEEbiographynophoto}{Eric Flamand}
got his PhD in Computer Science from INPG, France, in 1982. For the first part of his career he worked as a researcher with CNET and CNRS in France. He then held different technical management positions in the semiconductor industry, first with Motorola where he was involved into the architecture definition of the StarCore DSP. Then with ST microelectronics where he was first in charge of software development for the Nomadik Application Processor and then of the P2012 initiative aiming for a many core device. He is now co founder and CTO of Greenwaves Technologies a startup developing an IoT processor derived from PULP.
He is acting as a part time consultant for ETH Z\"urich.
\end{IEEEbiographynophoto}

\begin{IEEEbiographynophoto}{Frank K. G\"urkaynak}
has obtained his B.Sc. and M.Sc, in electrical engineering from the Istanbul Technical University, and his Ph.D. in electrical engineering from ETH Z\"urich in 2006. He is currently working as a senior researcher at the Integrated Systems Laboratory of ETH Z\"urich. His research interests include digital low-power design and cryptographic hardware.
\end{IEEEbiographynophoto}

\begin{IEEEbiographynophoto}{Luca Benini}
is the Chair of Digital Circuits and Systems at ETH Z\"urich and a Full Professor at the University of Bologna. He has published more than 700 papers in peer-reviewed international journals and conferences, four books and several book chapters. He is a member of the Academia Europaea.
\end{IEEEbiographynophoto}

\end{document}

%% file: 00_abstract.tex
Endpoint devices for Internet-of-Things not only need to work under extremely tight power envelope of a few milliwatts, but also need to be flexible in their computing capabilities, from a few kOPS to GOPS. Near-threshold\,(NT) operation can achieve higher energy efficiency, and the performance scalability can be gained through parallelism.
In this paper we describe the design of an open-source RISC-V processor core specifically designed for NT operation in tightly coupled multi-core clusters. We introduce instruction-extensions and microarchitectural optimizations to increase the computational density and to minimize the pressure towards the shared memory hierarchy. 
For typical data-intensive sensor processing workloads the proposed core is on average 3.5$\times$ faster and 3.2$\times$ more energy-efficient, thanks to a smart L0 buffer to reduce cache access contentions and support for compressed instructions. SIMD extensions, such as dot-products, and a built-in L0 storage further reduce the shared memory accesses by 8$\times$ reducing contentions by 3.2$\times$. 
With four NT-optimized cores, the cluster is operational from 0.6\,V to 1.2\,V achieving a peak efficiency of 67\,MOPS/mW in a low-cost 65\,nm bulk CMOS technology. In a low power 28\,nm FDSOI process a peak efficiency of 193\,MOPS/mW\,(40\,MHz, 1\,mW) can be achieved.

%% file: 01_intro.tex
In the last decade we have been exposed to an increasing demand for small, and battery-powered IoT endpoint devices that are controlled by a micro-controller\,(MCU), interact with the environment, and communicate over a low-power wireless channel. Such devices require ultra-low-power\,(ULP) circuits which interact with sensors. It is expected that the demand for sensors and processing platforms in the IoT-segment will skyrocket over the next years\,\cite{Lammel2015}.
Current IoT endpoint devices integrate multiple sensors, allowing for sensor fusion, and are built around a MCU which is mainly used for controlling and light-weight processing. 
Since endpoint devices are often untethered, they must be very inexpensive to maintain and operate, which requires ultra-low power operation. In addition, such devices should be scalable in performance and energy efficiency because bandwidth requirements vary from ECG sensors to cameras, to microphone arrays and so does the required processing power.
As the power of wireless (and wired) communication from the endpoint to the higher level nodes in the IoT hierarchy is still dominating the overall power budget\,\cite{Shnayder2004}, it is highly desirable to reduce the amount of transmitted data by doing more complex near-sensor processing such as feature extraction, recognition, or classifications\,\cite{Nakamura2007}. A simple MCU is very efficient for controlling purposes and light-weight processing, but not powerful nor efficient enough to run more complex algorithms on parallel sensor data streams\,\cite{konijnenburg2016}.

One approach to achieve a higher energy efficiency and performance is to equip the MCU with digital signal processing\,(DSP) engines which allow to e.g. extract the heart rate of an ECG signal more efficiently and reduce the transmission costs by only transmitting the extracted feature\,\cite{zhang2012}. Such DSPs achieve a very high performance when processing data, but are not as flexible as a processor and also harder to program.
An even higher energy efficiency can be achieved with dedicated accelerators. In a biomedical platform for seizure detection it has been shown that it is possible to speed up Fast Fourier Transformations\,(FFT) by a dedicated hardware block which is controlled by a MCU\,\cite{Sridhara2011}. Such a combination of MCU and FFT-accelerator is superior in performance, but also very specialized and hence, not very flexible nor scalable.

The question arises if it is possible to build a flexible, scalable and energy-efficient platform with programmable cores consuming only a couple of milliWatts. We claim that, although the energy efficiency of a custom hardware block can never be achieved with programmable cores, it is possible to build a flexible and scalable multi-core platform with a very high energy efficiency.
ULP-operations can be achieved by exploiting the near-threshold voltage regime where transistors become more energy-efficient\,\cite{Dreslinski2010}. The loss in performance\,(frequency) can be compensated by exploiting parallel computing. Such systems can outperform single core equivalents due to the fact that they can operate at a lower supply voltage to achieve the same throughput\,\cite{Dogan2012}.

A major challenge in low-power multi-core design is the memory hierarchy. Low-power MCUs typically fetch data and instructions from single-ported dedicated  memories. Such a simple memory configuration is not adequate for a multi-core system, but on the other hand, complex multi-core cache hierarchies are not compatible with extremely tight power budgets. Scratchpad memories offer a good alternative to data caches as they are smaller and cheaper to access\,\cite{banakar2002}. Another advantage is that such tightly-coupled-data-memories\,(TCDM) can be shared in a multi-core system and allow the cores to work on the same data structure without coherency hardware overhead.
One limiting factor in decreasing the supply voltage are memories which typically start failing first. The introduction of standard-cell-memories\,(SCMs) on the other hand allows for near-threshold operation and consume fewer milliWatts at the price of additional area\,\cite{meinerzhagen2010}.
In any case, memory access time and energy is a major concern in the design of a processor pipeline optimized for integration in an ULP multi-core cluster.

An open source ISA is a desirable starting point for an IoT core, as it can potentially decrease dependency from a single IP provider and cut cost, while at the same time allowing freedom for application-specific instruction extensions. Therefore, we focus on building a micro-architecture based on the \riscv instruction set architecture\,(ISA)\,\cite{Waterman2011} which achieves similar performance and code density to state-of-the art MCUs based on a proprietary ISA, such as ARM Cortex M series cores.
The focus of this work is on ISA and micro-architecture optimization specifically targeting near-threshold parallel operation, when cores are embedded in a tightly-coupled shared-memory cluster. Our main contributions can be summarized as follows:
\begin{itemize}
\item An optimized instruction fetch micro-architecture, featuring an L0 buffer with prefetch capability and support for hardware loop handling, which significantly decreases bandwidth and power in the instruction memory hierarchy.
\item An optimized execution stage supporting flexible fixed-point and saturated arithmetic operations as well as SIMD extensions, including dot-product and shuffle instructions, and misaligned load support that greatly reduce the load-store traffic to data memory while maximizing computational efficiency.
\item An optimized pipeline architecture and logic, RTL design and synthesis flow which minimize redundant switching activity and expose one full cycle time for accessing instruction caches and two cycles for the shared data TCDM, thereby giving ample timing budget to memory access that is most critical in near-threshold shared-memory operation.
\end{itemize}

The backend of the \riscv GCC compiler has been extended with fixed-point support, hardware loops, post-increment addressing modes and SIMD instructions\footnote{Both the RTL HW description and the GCC compiler are open source and can be downloaded at http://www.pulp-platform.org}. We show that with the help of the additional instructions and micro-architectural enhancements, signal processing kernels, such as filters, convolutions, etc. can be completed 3.5$\times$ faster on average, leading to a 3.2$\times$ higher energy efficiency. We also show that convolutions can be optimized in C with intrinsics by using the dot-product and shuffle instruction, which reduces accesses and contentions for the shared data memory utilizing the register file as L0-storage allowing the system to achieve a near linear speedup of 3.9$\times$ when using four cores instead of one.

The following sections will first summarize the related work in \secref{sec:relwork} and explain the target multi-core platform and its applications in \secref{sec:pulp}. The core architecture is presented in \secref{sec:core} with programming examples in \secref{sec:sw}. Finally, \secref{sec:results} discusses the experimental results and in \secref{sec:conclusion} we draw our conclusions.

%% file: 02_relwork.tex
The majority of IoT endpoint devices use single core MCUs for controlling and light-weight processing. Single-issue in-order cores with a high IPC are typically more energy-efficient as no operations have to be repeated due to mispredictions and speculation\,\cite{Azizi2010}. Commercial products often use ARM processors from the Cortex-M families~\cite{STM32,Ambiqmicro2015,NXP2016} which work above the threshold voltage. Smart peripheral control, power managers and combination with non-volatile memories allow these systems to consume only tens of milliWatts in active state, and a few microWatts in sleep mode. In this work we will focus on active power and energy minimization. Idle and sleep reduction techniques are surveyed in\,\cite{SiddharthRele2002}.
\\
Several MCUs in the academic domain and a few commercial ones exploit the use of near-threshold operation to achieve energy-efficiencies in active state down to 10\,pJ/op\,\cite{ickes2011,Ambiqmicro2015}. These designs take advantage of standard cells and memories which are functional at very low voltages. It is even possible to operate in the sub-threshold regime and achieve excellent energy-efficiencies of 2.6\,pJ/op at 320\,mV and 833\,kHz\,\cite{Zhai2009}. Such systems can consume well below 1mW active power, but reach their limit when more than a few MOPS of computing power is required as for near-sensor processing for IoT endpoint devices.

One way to increase performance while maintaining power in the mW range is to use DSPs which make use of several optimizations for data intensive kernels such as parallelism through very long instruction words\,(VLIW) and specialized memory architectures. Low-power VLIW DSPs operating in the range of 3.6-587\,MHz where the chip consumes 720\,\textmu W to 113\,mW have been proposed\,\cite{gammie2011}. It is also possible to achieve energy-efficiencies in the range of a couple of pJ/op. Wilson et al. designed a VLIW DSP for embedded $F_{max}$ tracking with only 62\,pJ/op at 0.53V\,\cite{wilson2014460mhz}, and a 16b low-power fixed-point DSP with only about 5\,pJ/op has been proposed by Le et al.\,\cite{Le2015}. DSPs typically have zero overhead loops to eliminate branch overheads and can execute operations in parallel, but are harder to program than general purpose processors. In this work we borrow several ideas from the DSP domain, but we still maintain complete compatibility with the streamlined and clean load-store RISC-V ISA, and full C-compiler support (no-assembly level optimization needed) with a simple in-order four-stage pipeline with Harvard memory access.

Since typical sensor data from ADCs uses 16b or less, there is a major trend to support SIMD operations in programmable cores such as in the ARM Cortex M4\,\cite{CortexM4} which supports DSP functionalities while remaining energy-efficient (32.8\,\textmu W/MHz in 90\,nm low power technology\,\cite{CortexM4}).
The instruction set contains DSP instructions which offer a higher throughput as multiple data elements can be processed in parallel. ARM even provides a Cortex-M Software Interface Standard\,(CMSIS) DSP library which contains several functions which are optimized with builtins\,\cite{CortexM4}. Performance can for example be increased with a dot-product instruction which accumulates two 16b$\times$16b multiplication results in a single cycle.
Such dot-product operations are suitable for media processing applications\,\cite{Farooqui2000} and even versions with 8b inputs can be implemented\,\cite{Zhang2009} and lead to high speedups. While 16b dot-products are supported by ARM, the 8b equivalent is not.

Dedicated hardware accelerators, coupled with a simple processor for control, can be used for specific tasks offering the ultimate energy efficiency. As an example, a battery-powered multi-sensor acquisition system which is controlled by an ARM Cortex M0 and contains hardware accelerators for heart rate detection has been proposed by Konijnenburg et al.\,\cite{konijnenburg2016}.
Ickes et al. propose another system with FFT and FIR filter accelerators which are controlled by a 16b CPU and achieves an energy efficiency of 10\,pJ/op at 0.45V\,\cite{ickes2008}.
Also convolutional engines\,\cite{Qadeer2013,Cavigelli2016,Conti2015} which outperform  general purpose architectures have been reported. Hardwired accelerators are great to speedup certain tasks and can be a great improvement to a general purpose system as described in\,\cite{Conti2015}. However, since there is no standard to interface such accelerators, and the fact that such systems cannot be scaled up, prompts us to explore in this paper a scalable multi-core platform which covers all kind of processing demands and is fully programmable.

NXP, TI and other vendors offer heterogenous dual-core MCUs featuring decoupled execution of an extremely energy-efficient Cortex M0+ for control tasks, and a Cortex M3 or M4 for computationally-intensive kernels~\cite{NXP2016,TICC2650}. Such systems are not scalable in terms of memory hierarchy and number of cores as cores are embedded in their own subsystems and M0+ cannot run M3/4 executables.
A few multi-core platforms have already been proposed. E.g. Neves et al. proposed a multi-core system with SIMD capabilities for biomedical sensors~\cite{Neves2015} and Hsu et al. implemented a reconfigurable SIMD vector machine which is very powerful, energy-efficient and scalable in performance\,\cite{Hsu2012}. Both are specialized processors which are not general purpose, but optimized for a specific domain and very difficult to program efficiently in C.
A very powerful multi-core architecture consisting of 64 cores operating near the threshold voltage has been proposed by\,\cite{Fick2013}. The system is designed for high performance computing and its power consumption is in the range of a couple of hundreds milliWatts and not suitable for IoT-applications.

%% file: 03_pulp.tex
In this section we briefly review the parallel ultra-low power (PULP) cluster architecture which embeds our \riscv cores. Interested readers are refered to\,\cite{Rossi2016,Rossi2016a,Pullini2016} for more details. Our focus is on highlighting the key elements of the PULP memory hierarchy and the targeted workloads, which have been the main drivers of the core design. Many MCUs operate on instruction, and data memories and do not use caches. 

\begin{figure}[tb]
\centering
  \includegraphics[width=0.95\linewidth]{\figpath/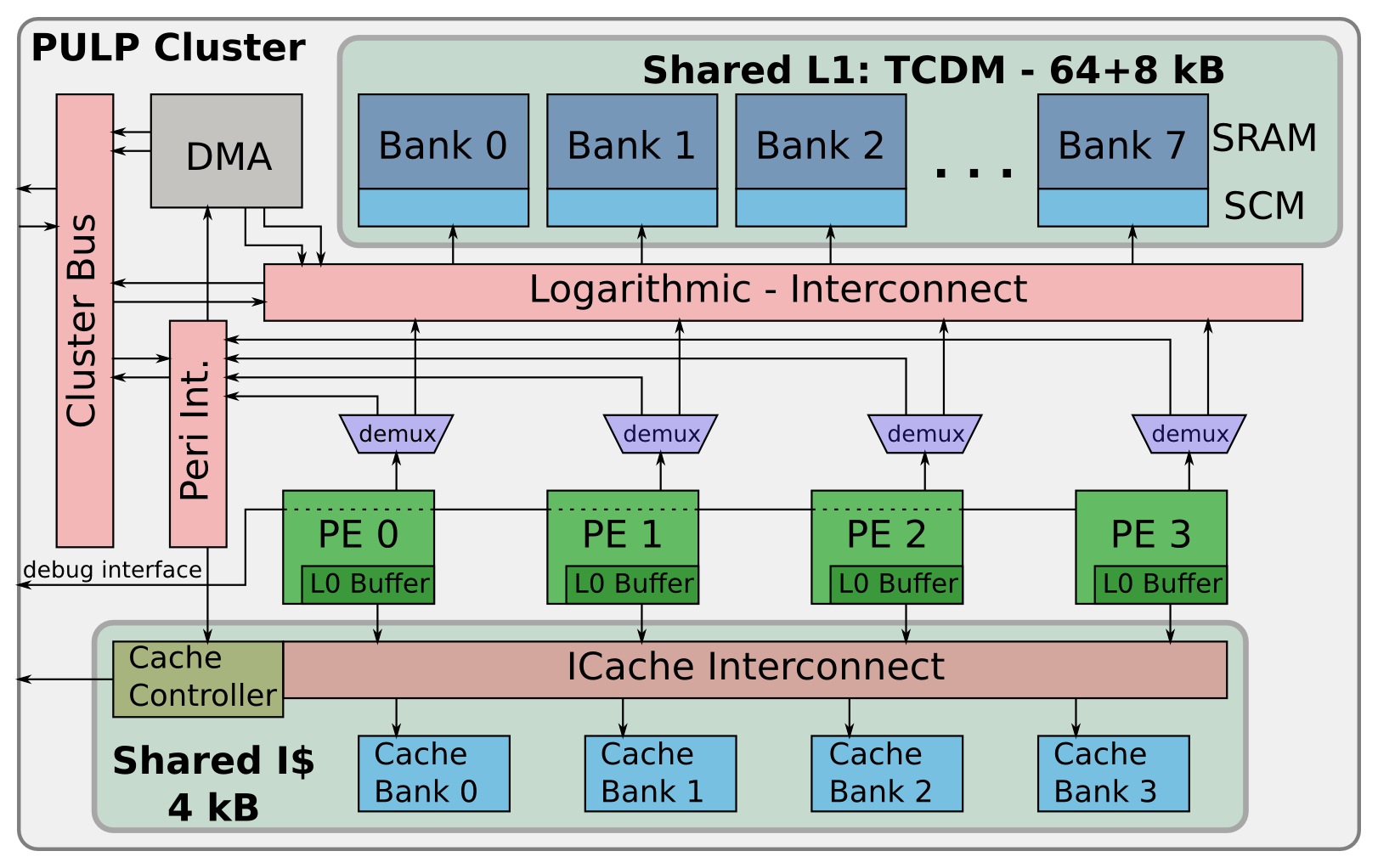}
  \caption{The PULP cluster with four cores, eight shared TCDM-banks of 1\,kB SCM and 8\,kB SRAM each and a shared I\$ of 4\,kB used for evaluations in this paper. The cores (PE) are \riscv architectures with extended DSP capabilities.}
  \label{fig:pulp}
\end{figure}

Parallel execution in the PULP cluster requires a scalable memory architecture to achieve near-ideal speedups in parallel applications, while curtailing power. A PULP cluster supports a configurable number of cores with a shared instruction cache and scratchpad memory. Both memory ports have variable latency and can stall the pipeline.
\figref{fig:pulp} shows a PULP cluster in a configuration with 4 cores, and 8 TCDM-banks. A shared I\$ is used to reduce the cost per core, leveraging the single-program-multiple-data nature of most parallel near-sensor processing application kernels. A tightly coupled DMA engine manages transfers between IO and L2 memory and the shared TCDM. Data access pattern predictability of key application kernels, the relatively low clock frequency target and the tightly constrained area and power budget make a shared TCDM preferable over a collection of coherent private caches\,\cite{banakar2002}.

The data memories are split in area-efficient per-core SRAM, and energy-efficient SCM blocks. Since SCMs are built of standard cells, it is possible to scale the supply voltage and operate near the threshold voltage of transistors\,\cite{meinerzhagen2010}. By clock-gating the SRAM blocks and the individual cores, it is therefore possible to scale down the system to a simple single core architecture which can act as an extremely energy-efficient controller. If more processing power is required, the power manager of the cluster can wake up more cores and by dynamic-voltage and frequency-scaling\,(DVFS), the performance of a 28nm FDSOI implementation can be adjusted from a couple of operations per second to 2\,GOPS by scaling the voltage from 0.32V to 1.15V where the cores run at 500\,MHz.
The PULP-cluster has been successfully taped out with OpenRISC, and \riscv cores\,\cite{Rossi2016,Rossi2016a,Pullini2016} and its latest version achieves an energy efficiency of 193 MOps/mW in 28\,nm FDSOI technology.

In this paper, we will focus on ISA extensions, micro-architecture optimizations and RTL design to further improve the energy efficiency of the RISC-V core used in PULP. It is well known that memory accesses for both data and instructions are the most critical operations that contribute to energy consumption in a microprocessor as we will show in \secref{sec:instruction}, and we will present several methods to reduce costly memory access operations for both data and instructions. 

The \riscv ISA used as a starting point in this paper already supports a compressed ISA-extension that allows several common instructions to be coded in 16b instead of 32b, reducing the pressure on the instruction cache (I\$). The PULP architecture is designed to use a shared I\$ to decrease the per-core instruction fetch cost. A shared I\$ is especially efficient when the cores execute the same parallel program (e.g. when using OpenMP). 
Since PULP, unlike GP-GPUs, does not enforce strict single-instruction-multiple data execution, the shared cache may produce stalls that add to energy losses.
We have therefore added a L0-buffer into the core to reduce access contentions, reduce instruction read power, and shorten the critical path between I\$ and the core. In addition we have modified the L0-buffer to cope with non-aligned accesses due to compressed instructions more efficiently as discussed in \secref{sec:if}.
As for the data access, we will illustrate the major gains that can be achieved by our extensions using the example of a 2D-convolution implementation that is widely used in many applications in the image processing domain\,\cite{Sonka1999}. 2D-convolutions are not only very pervasive operators in state-of-the-art sensor data processing\footnote{For instance convolution make more than 80\% of the workload in Deep Convolutional Neural Networks}, but they also contain the same iterative patterns as digital filtering (FIR, IIR) and statistical signal processing (matrix-vector product) operations. The main task of a convolution operation is to multiply subsequent values with a set of weights and accumulate all values for which a Multiply-Accumulate (MAC) units would be most commonly used. 

\begin{figure}[h]
\centering
  \includegraphics[width=0.9\linewidth]{\figpath/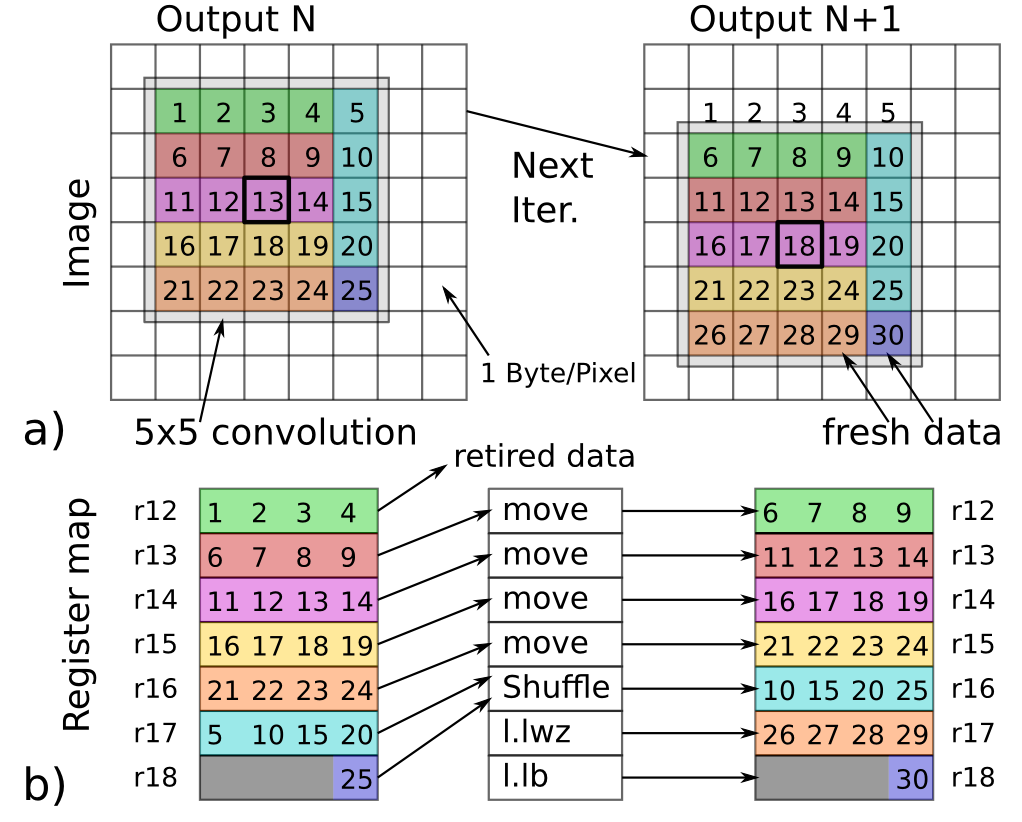}
  \caption{a) Example of a 5$\times$5 convolution to compute output N and N+1 in the image domain and b) how the register content is efficiently updated using the shuffle instruction. One 5$\times$5 convolution requires exactly 4 \emph{move}, one \emph{shuffle}, and 2 loads to prepare the operands and 1 \emph{dotp}-, and 6 \emph{sdotp}-operations to compute one output pixel.}
  \label{fig:conv}
\end{figure}

A straightforward implementation of a 5$\times$5 2D-convolution of a 64$\times$64 pixel image on the \riscv architecture with compressed instructions takes 625k cycles, of which 389k are for various load/store (\emph{ld/st}) operations to and from memory and 100k are actual operations calculating the values.
The same algorithm when parallelized to four cores on the PULP requires only 173k cycles per convolution on average. However, increased parallelism also results in increased demand to the shared TCDM which puts additional strain on the TCDM design. TCDM-contentions can be reduced by increasing the number of banks, which unfortunately in turn increases power-per-access due to more complex interconnection and increased memory area because smaller and less dense memories have to be used. Improving the arithmetic capabilities of the cores will further exacerbate this problem as it will lead to a higher portion of \emph{ld/st} instructions which will result in even more contentions (up to 20\%).

A common technique to reduce the data bandwidth is to make use of vector instructions that process instead of a single 32\,bit value, two 16\,bit values, or four 8\,bit values simultaneously. Most arithmetic and logic operations can be parallelized in sub-fields with very little overhead. As explained in more detail in \secref{sec:core}, we have added a large variety of vector instructions to the \riscv ISA including dot product (\emph{dotp, sdotp}) instructions that can multiply four pairs of 8\,bit numbers and accumulate them in a single instruction. 

Allowing one data word to hold multiple values directly reduces the memory bandwidth for data accesses, which is only useful if the data at sub-word level can be efficiently manipulated without additional instructions. \figref{fig:conv} explains how a 5$\times$5 2D-convolution can be computed with vector instructions. It can be seen that for each convolution step, 25 data values have to be multiplied with 25 constants and accumulated. If 8\,bit values are being used, registers can be used to hold vectors of four elements each. Once this calculation is completed, for the next step of the iteration, the five values of the first row will be discarded, and a new row of five values will be read. If these vectors are not aligned to word boundaries, an unaligned word has to be loaded from memory which can be supported either in hardware or software. A software implementation requires at least five instructions to load two words and combine the pixels in a vector. In addition, it blocks registers from being used for actual computations, which is the reason why we support unaligned memory accesses directly in the load-store-unit\,(LSU) by issuing two subsequent requests to the shared memory. Hence, unaligned words can be loaded in only two cycles. We also implement the \emph{shuffle} instruction that can combine sub-words from two registers in any combination. 
\figref{fig:conv}b) shows how \emph{move} and \emph{shuffle} instructions are used to recombine the pixels in the right registers instead of loading all elements from memory. This allows to reduce register file pressure and the number of loads per iteration from 5 to 2.
One iteration can therefore be computed in about 20~instructions (4 \emph{move}, 1 \emph{shuffle}, 2 \emph{load}, 7 \emph{dotp}, 1 \emph{store}, 5 control flow), or 26 cycles on average. Thus, when all of the improvements are combined, the time to complete the operations can be reduced by 14.72$\times$ when compared to the original single core implementation. Coupled by efficient DVFS methods used in a PULP cluster this performance gain can be used to increase energy efficiency by working at NT-operation or to reduce computation time at the same operation voltage, allowing the system a wide range of tunability which will be very important for future IoT systems that need to adapt to a variety of operating conditions and computation loads.

%% file: 04_core_archi.tex
In this section, we will detail the extensions made to the \riscv ISA and micro-architectural optimizations for increasing the efficiency of the processor core when working in a multi-core platform with shared memory. The pipeline architecture will be described first in \secref{sec:pipe} and the individual components of the core are discussed starting with \secref{sec:if} where the IF-stage with the pre-fetch buffer to support compressed instruction is explained. Hardware-loop and post-increment extensions to the \riscv ISA are explained in \secref{sec:hwloop}, and \ref{sec:postmod}. The EX-stage with a more powerful dot-product-multiplier and a vector-ALU with fixed-point support and a shuffle-unit will be explained in \secref{sec:alu} and \ref{sec:mult}.

\subsection{Pipeline Architecture}
\label{sec:pipe}

\begin{figure*}[tb]
 \centering
    \includegraphics[width=0.9\linewidth]{\figpath/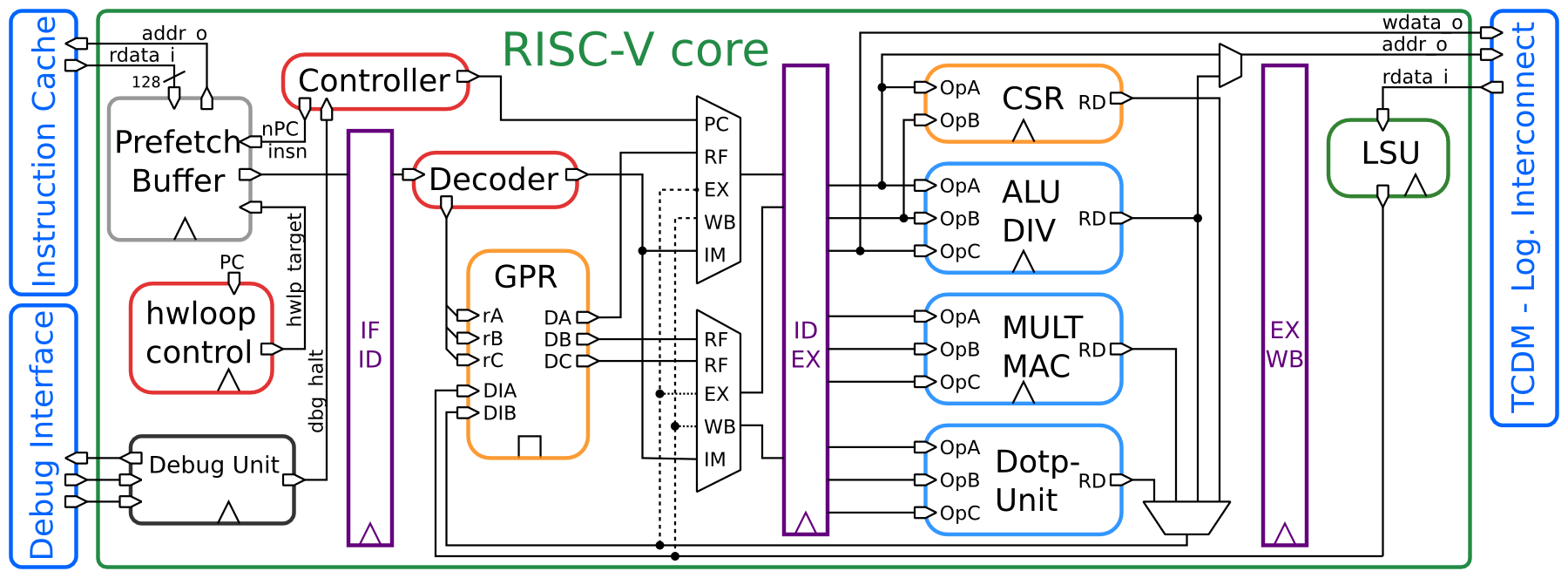}
    \caption{Simplified block diagram of the \riscv core architecture showing its four pipeline stages and all functional blocks.}
    \label{fig:core_archi}
\end{figure*}

The number of pipeline stages used to implement the processor core is one of the key design decisions. A higher number of pipeline stages allows for higher operating frequencies, increasing overall throughput, but also increases data and control hazards, which in turn reduces the IPC. For high-performance processors, where fast operation is crucial,  optimizations such as branch predictions, pre-fetch-buffers, and speculation can be introduced. However, these optimizations add to the overall power consumption and are usually not viable for ultra-low power operation where typically a shallow pipeline of 1-5 stages with a high IPC is preferred. 

The basic goal to keep the IPC high is to reduce stalls as much as possible. The ARM Cortex M4, for example, features a 3-stage fetch-decode-execute pipeline with a single write-back port on the register file\,(RF). In this architecture, the absence of a separate write-back port for the load-store unit\,(LSU) leads to stalls whenever an ALU-operation follows a load-operation. ARM solves this issue by grouping load-words operations as much as possible to avoid stalls. In a multi-core environment with shared memory, such as PULP, this optimization will increase the number of access contentions at the shared memory. This is why we have opted for a 4-stage pipeline with a separate write-back port on the RF for the LSU. Multi-ported RFs are more expensive than single ported ones, but can be implemented with low power overhead\,\cite{Zeng2015}. The additional write-port also eliminates the write-back muxes (ALU, CSR, MULT) from the critical path while increasing the area of the register file by only 1\,kGE. The organization of the pipeline and all functional units are shown in \figref{fig:core_archi}. The pipeline is organized in four stages, \emph{instruction fetch}\,(IF), \emph{instruction decode}\,(ID), \emph{execute}\,(EX), and \emph{write-back}\,(WB).

The TCDM in the multi-core PULP architecture is organized in multiple banks and can be accessed over a logarithmic interconnect. Arbitration and the interconnect add delay to the data request and return paths. The 4-stage pipeline organization also allowed us to better balance the request and return paths from the shared memory banks. In practical implementations, we have employed \emph{useful skew} techniques to balance the longer return path from the memory by skewing the clock of the core\footnote{In a 65\,nm technology implementation with 2.8\,ns clock period for 1.08\, worst-case conditions, 0.5\,ns useful skew was employed.}. The amount of skew depends on the available memory macros, the number of cores and memory banks in the cluster.
Even though the data interface is limiting the frequency of the cluster, the cluster is ultimately achieving frequencies of 350-400\,MHz when implemented in 65\,nm under typical conditions, and the PULP cluster reaches higher frequencies than commercially available MCUs operating in the range of 200\,MHz. Since the critical path is mainly determined by the memory interface, it was possible to extend the ALU with fixed-point arithmetic and a more capable multiplier that supports \emph{dotp} operations  without incurring additional timing penalties. 

\subsection{Instruction Fetch Unit}
\label{sec:if}

Similar to the Thumb-2 instruction set of ARM, the \riscv standard contains a specification for compressed instructions which are only 16b long and mirror full instructions with a reduced immediate size and RF-address. As long as these 16b instructions are aligned to 32b word boundaries, they are easy to handle. A compressed instruction decoder unit detects and decompresses these instructions into standard 32b instructions, and stalls the fetch unit for one cycle whenever 2 compressed instructions have been fetched.

\begin{figure}[!b]
 \centering
    \includegraphics[width=0.85\linewidth]{\figpath/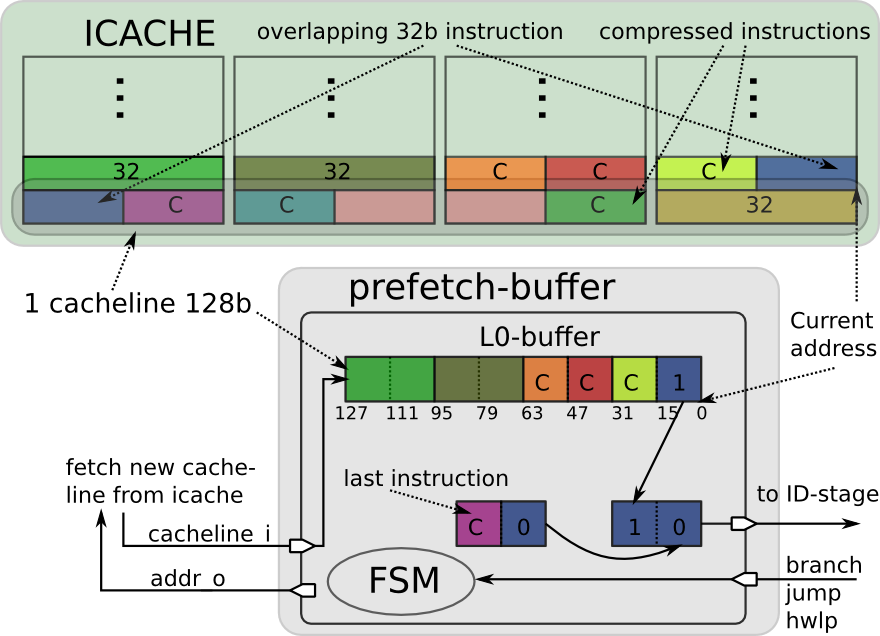}
    \caption{Block Diagram of the pre-fetch buffer with an example when fetching full instructions over cache-line boundaries.}
    \label{fig:prefetch}
\end{figure}

Inevitably some 32b instructions will become unaligned when an odd number of 16b instructions are followed by a 32b instruction, requiring an additional fetch cycle for which the processor will have to be stalled. As described earlier in \secref{sec:pulp}, we have added a pre-fetch-buffer which allows to fetch a complete cache line (128b) instead of a single instruction to reduce the access contentions associated with the shared I\$\,\cite{Loi2015}. While this allows the core to access 4 to 8 instructions in the pre-fetch-buffer, the problem of fetching unaligned instructions remains, it is just shifted to cache-line boundaries as illustrated in \figref{fig:prefetch} where the current 32b instruction is split over two cache-lines. 

To prevent the core from stalling in such cases, an additional register is used to keep the last instruction. In the case of an unaligned 32b instruction, this register will contain the lower 16b of the instruction which can be combined with the higher part\,(1) and forwarded to the ID-stage. This addition allows unaligned accesses to the I\$ without stalls unless a branch, hardware loop, or jump is processed. In these cases, the FSM has to fetch a new cache-line to get the new instruction. The area cost of this pre-fetch buffer is mainly due to additional registers and accounts for 4\,kGE.

\input{\tabpath/tabisa.tex}

\subsection{Hardware-loops}
\label{sec:hwloop}
\emph{Zero-overhead loops} are a common feature in many processors, especially DSPs, where a hardware loop-controller inside the core can be programmed by the loop count, the beginning and end-address of a loop. Whenever the current program counter (PC) matches the end-address of a loop, and as long as the loop counter has not reached 0, the hardware loop-controller provides the start-address to the fetch engine to re-execute the loop. This eliminates instructions to test the loop counter and perform branches, thereby reducing the number of instructions fetched from I\$. 

The impact of hardware loops can be amplified by the presence of a loop buffer, i.e. a specialized cache holding the loop instructions, which removes any fetch delay\,\cite{Uh2000}, in addition fetch power can also be reduced by the presence of a small loop cache\,\cite{Bajwa1997}. Nested loops can be supported with multiple sets of hardware loops, where the innermost loop always gives the highest speedup as it is the most frequently executed. We have added hardware loop support to the \riscv cores at the micro-architectural level with only two additional blocks: A controller and a set of registers to store the loop information. Each set has associated 3 special registers to hold the loop counter, the start- and the end-address. The registers are mapped in the CSR-space which allows to save and restore loop information when processing interrupts or exceptions. A set of dedicated instructions have been provided that will initialize a hardware loop in a single instruction using \emph{lp.setup} (or \emph{lp.setupi}). Additional instructions are provided to set individual registers explicitly (\emph{lp.start}, \emph{lp.end}, \emph{lp.count}, \emph{lp.counti}). 

Since the performance gain is maximized when the loop body is small, supporting many register sets only brings marginal performance improvements at a non-negligible cost in terms of area ($\approx$ 1.5\,kGE per register set). Our experiments have shown that two register sets provide the best trade-off. As mentioned earlier, our improved cores feature a pre-fetch-buffer able to hold 4 to 8 instructions of the I\$. This pre-fetch-buffer can act as a very small loop cache and if the loop body fits into the pre-fetch-buffer, I\$ accesses can be eliminated during the loop, reducing the power considerably. The GCC compiler has been modified to automatically insert hardware loops by using the dedicated instructions provided in \tabref{tab:newinstructions}.

\subsection{Load-store unit}
\label{sec:postmod}
The basic \riscv architecture only supports one type of \emph{ld/st} operation where the effective address is computed by adding an offset coming from an immediate to the base address stored in a register. We have first added an additional addressing mode where the offset can be stored in a register instead of an immediate, and then added a post-increment addressing mode with an immediate or register offset to automatically update pointers. A pre-increment \emph{ld/st} mode was not deemed necessary as every pre-increment \emph{ld/st} operation can be rewritten in a post-increment \emph{ld/st} operation. Support for post-increment instructions leads to high speedup of up to 20\% when memory access patterns are regular as it is the case for example in a matrix multiplication. To support \emph{ld} operations with post-increment, two registers have to be updated in the RF: the data from memory and the incremented address pointer which is computed in the ALU. Since ALU and LSU have separate register file write ports, both values can be written back without experiencing any contentions. \tabref{tab:newinstructions} shows the various forms of additional \emph{ld/st}-instructions.

\begin{figure}[tb]
 \centering
    \includegraphics[width=0.56\linewidth]{\figpath/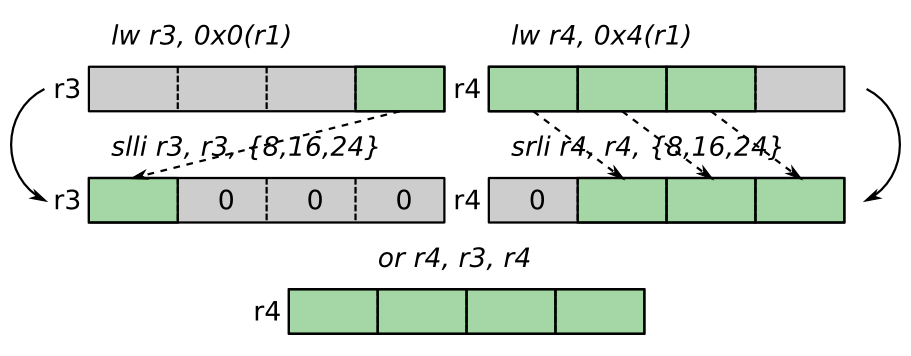}
    \includegraphics[width=0.4\linewidth]{\figpath/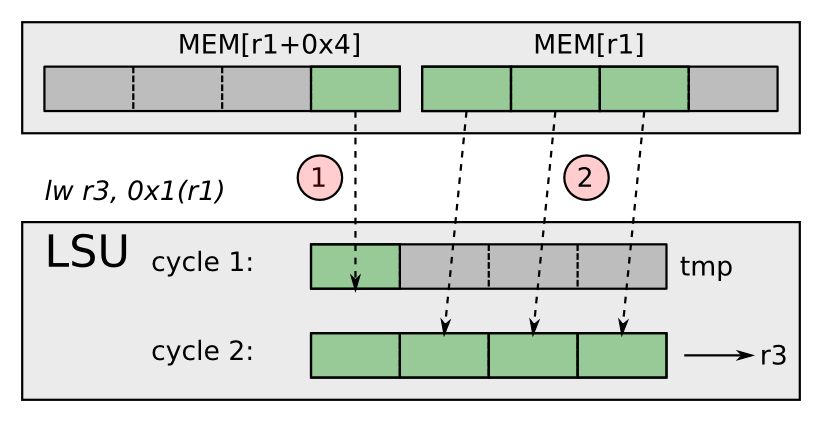}
    \caption{a) support for unaligned access in software (5 instructions/cycles) and b) with hardware support in the LSU (1 instruction, 2 cycles).}
    \label{fig:unaligned}
\end{figure}

The LSU has also been modified to support unaligned data memory accesses which frequently happen during vector operations such as the 5$\times$5 2D-convolution described earlier in \secref{sec:pulp}. If the LSU detects an unaligned data access, it issues first a request to the high word and stores the data in a temporary register. In a second request, the lower bytes are accessed, which are then on the fly combined with the temporary register and forwarded to the RF. This approach not only allows to reduce the pressure on the number of used registers, but also reduces code size as shown in \figref{fig:unaligned}, and the number of required cycles to access unaligned vector elements. In addition, this low-overhead approach is better than full hardware support as this would imply to double the width of the interconnect and change to the memory architecture.

\subsection{EX-Stage: ALU}
\label{sec:alu}
\subsubsection{Packed-SIMD support}
\label{sec:packedsimd}
To take advantage of applications in the IoT domain that can work with 8b and 16b sensor data, the ALU of a 32b microprocessor can be modified to work on vectors of four and two elements using a vectorized datapath segmented into two or four parts, allowing to compute up to four bytes in parallel. Such operations are also known as subword parallelism~\cite{Lee1996}, packed-SIMD or micro-SIMD~\cite{Shahbahrami2004} instructions.

We have extended the RVC32IM ISA with sub-word parallelism for 16b (halfword) and 8b (byte) operations in three addressing variations. The first variation uses two registers, the second uses an immediate value and the third replicates the scalar value in a register as the second operand for the vectorial operation. Vectorial operations like additions or subtractions have been realized by splitting the operation in four sub-operations which are connected through the carry propagation signals. For example, the full 32b result is computed by enabling all the carry propagation signals while in vector addition at byte level, the carry propagation is terminated between sub-operations. The final vectorial adder is therefore a 36b architecture (32b for operands, and 4 carry bits). Vectorial comparison and shift operations have similarly been realized by splitting the datapath in four separate segments. A 32b comparison is then computed using the four 8b comparison results. Logic vectorial operations as \emph{and, or} and \emph{xor} are trivial since the exact same datapath can be used.

\begin{figure}[h]
\centering
  \includegraphics[width=0.7\linewidth]{\figpath/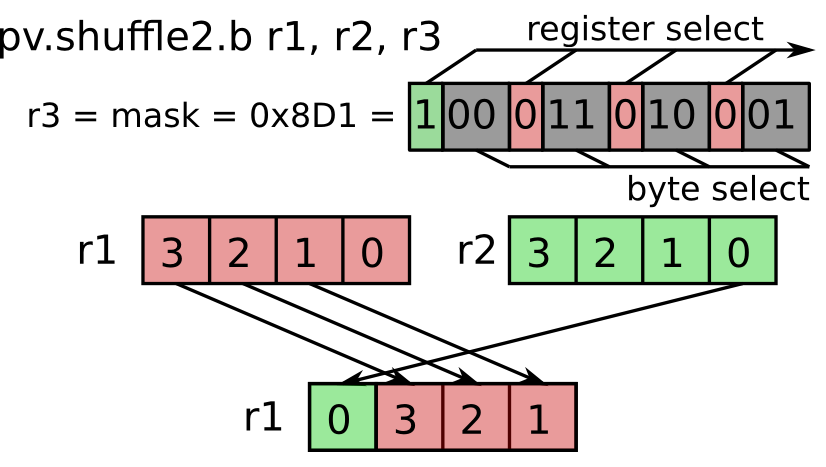}
  \caption{The Shuffle instruction allows to efficiently combine 8b, or 16b elements of two vectors in a single one. For each byte the mask encodes which byte (index) is used from which register (select).}
  \label{fig:shuffle}
\end{figure}

Additional sub-word data manipulation instructions are needed to prepare vector operands\,\cite{Chang2006} for vector ALUs. We have implemented a three operand \emph{shuffle} instruction that can generate the output as any combination of the sub-words of two input operands, while the third operand, sets selection criteria either through an immediate value or a register as seen in \figref{fig:shuffle}. The \emph{shuffle} instruction is supported through a tree of shared multiplexers and can also be used to implement further sub-word data manipulation instructions such as: \emph{insert} to overwrite only one sub-word, \emph{extract} to read only one sub-word and \emph{pack} to combine sub-words of two registers in single vector.

\subsubsection{Fixed-Point support}
\label{sec:alufixed}

\begin{figure}[tb]
 \centering
    \includegraphics[width=0.85\linewidth]{\figpath/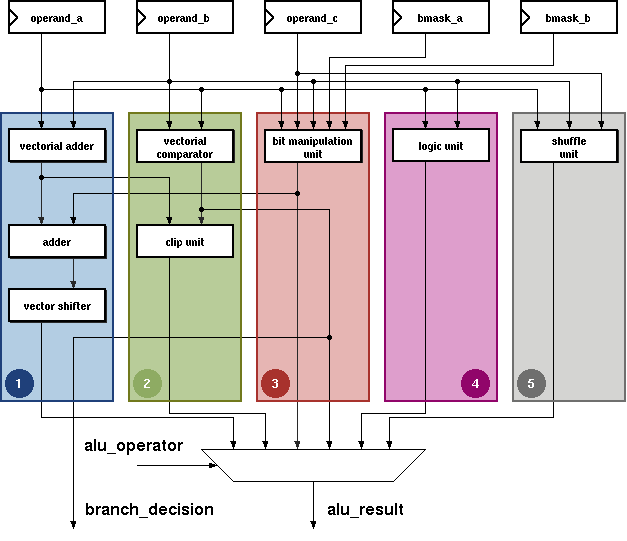}
    \caption{Simplified block diagram of the \riscv ALU.}
    \label{fig:alu_archi}
\end{figure}

There are many applications, such as  speech processing, where floating-point accuracy is not always needed, and simpler fixed-point arithmetic operations can be used instead\,\cite{Chang2016}. This has the advantage of re-using the integer datapath for most operations with the help of some dedicated instructions for fixed-point support, such as saturation, normalization and rounding. 

Fixed-point numbers are often given in the Q-Format where a Q$n$.$m$ number consists of $n$ integer bits and $m$ fractional bits. Some processors support a set of fixed-point numbers encoded in 8b, 16b or 32b, and provide dedicated instructions to handle operations with these numbers. For example, the ARM Cortex M4 ISA provides instructions as QADD or QSUB to add two numbers and then saturate the results to 8b, 16b or 32b.

\begin{table}[tb]
\small
\begin{threeparttable}
\renewcommand\arraystretch{0.3}
\caption{Addition of four Q1.11 fixed-point numbers with and w/o fixed-point instructions.}
\centering
\label{tab:addq111}
\begin{tabular*}{0.49\textwidth}{l@{\extracolsep{\fill}}*{2}{l}}
\toprule
Without Add Norm Round & With Add Norm Round \\
\midrule
  \texttt{add  ~r3, r4, r5} &   \texttt{add ~~~~r3, r4, r5}       \\
  \texttt{add  ~r3, r3, r6} &   \texttt{add ~~~~r3, r3, r6}       \\
  \texttt{add  ~r3, r3, r7} &   \texttt{p.addRN r3, r3, r7, 2} \\
  \texttt{addi r3, r3, 2}  &                         \\
  \texttt{srai r3, r3, 2} &                         \\
\bottomrule
\end{tabular*}
\end{threeparttable}
\end{table}

Our extensions to the \riscv architecture has been designed to support fixed-point arithmetic operations in any Q-format with the only limitation that $n+m < 32$. We provide instructions that can add/subtract numbers in fixed-point arithmetic and shift them by a constant amount to perform normalization. The two code examples in \tabref{tab:addq111} show how the combined add-round-normalize (\emph{p.addRN}) instruction can save both code size (3 instead of 5 instructions) and execution time (2 cycles less). In this example, four numbers represented by Q1.11 are summed up. The result, if not normalized will be a 14\,bit long Q3.11 number. To keep the result in 12\,bits, rounding can be achieved by adding 2 units of least precision (ulp) to the result and shift the number right by 2 places. The result can then be interpreted as a 12 bit number in Q3.9 format. The final \emph{p.addRN} instruction achieves this rounding in a single step by first adding the two operands using the vectorial adder, adding $2^{(I-1)}$ to the intermediate result before it is shifted by $I$ bits utilizing the shifter of the ALU. An additional 32b adder was added to the ALU to help with the rounding operation as seen in the highlighted region 1 of \figref{fig:alu_archi}.

For fixed point operations, a \emph{clip} instruction has been implemented to check if a number is between two values and saturates the result to a minimum or maximum bound otherwise. No significant hardware has been added to the ALU to implement the clip instruction, indeed the \textit{greater than} comparison is done using the existing comparator and the \textit{less than} comparison is done in parallel by the adder. Unlike the ARM Cortex M4 implementation, our implementation requires an additional \emph{clip} instruction, but has the added benefit of supporting any Q-number format and allows to round and normalize the value before saturating which provides higher precision.

\tabref{tab:clip} shows an example of compiler-generated code where two arrays, each containing \textit{n} Q1.11 signed elements are added together, then the result is normalized between $-1$ and $1$ represented in the same Q1.11 format. The example clearly illustrates the difference between the \riscv ISA with and without clip support. \tabref{tab:newinstructions} shows the added instructions for fixed-point support.
Note that the code to the right is not only shorter, it also does not have control flow instructions, thereby achieving better IPC.

\begin{table}[tb]
\small
\begin{threeparttable}
\renewcommand\arraystretch{0.3}
\caption{Addition of \textit{n} elements with saturation.}
\centering
\label{tab:clip}
\begin{tabular*}{0.49\textwidth}{l@{\extracolsep{\fill}}*{2}{l}}
\toprule
Without clip support & With clip support \\
\midrule
  \texttt{addi ~~~~~r15, r0, 0x800     }  & \texttt{addi ~~~~~r3, r0, \textit{n} }  \\
  \texttt{addi ~~~~~r14, r0, 0x7FF     }  & \texttt{lp.setupi r0, r3, endL  }  \\
  \texttt{addi ~~~~~r3, r0, \textit{n} }  & \texttt{p.lh   ~~~~~0(r10!), r4      }  \\
  \texttt{lp.setupi r0, r3, endL  }       & \texttt{p.lh  ~~~~~0(r11!), r5      }  \\
  \texttt{p.lh   ~~~~~0(r10!), r4      }  & \texttt{add  ~~~~~~r4, r4, r5         }  \\
  \texttt{p.lh   ~~~~~0(r11!), r5      }  & \texttt{p.clip ~~~r4, r4, 12       }  \\
  \texttt{add  ~~~~~~r4, r4, r5         }  & \texttt{endL: sw ~0(r12!), r4    }  \\
  \texttt{blt  ~~~~~~r4, r15, lb        }  & \texttt{                        }  \\
  \texttt{blt  ~~~~~~r14, r4, ub        }  & \texttt{                        }  \\
  \texttt{j    ~~~~~~~~endL             }  & \texttt{                        }  \\
  \texttt{lb: mv ~~~r4, r15         }  & \texttt{                        }  \\
  \texttt{j    ~~~~~~~~endL               }  & \texttt{                        }  \\
  \texttt{ub:  mv ~~~r4, r14         }  & \texttt{                        }  \\
  \texttt{endL: sw ~0(r12!), r4    }  & \texttt{                        }  \\
\bottomrule
\end{tabular*}
\end{threeparttable}
\end{table}

\subsubsection{Bit manipulation support}
There are many instances where a single bit of a word needs to be accessed 
e.g. to access a configuration bit of a memory-mapped register. We have enhanced the \riscv ISA with instructions such as \emph{p.extract} (read a register set of bits), \emph{p.insert} (write to a register set of bits), \emph{p.bclr, p.bset} (clear/set a set of bits), \emph{p.cnt} (count number of bits that are 1), \emph{p.ff1,p.fl1} (find index of first/last bit that is 1 in a register) and \emph{p/clb} (count leading bits in a register).

\subsubsection{Iterative Divider}
To fully support the RVC32IM \riscv specification we have opted to support division by using long integer division algorithm by reusing existing comparators, shifters and adders of the ALU. Depending on the input operands, the latency of the division operation can vary from 2 to 32 cycles. While it is slower than a dedicated parallel divider, this implementation has low area overhead (2\,kGE).

\subsection{EX-Stage: Multiplication}
\label{sec:mult}

\begin{figure}[tb]
 \centering
    \includegraphics[width=0.85\linewidth]{\figpath/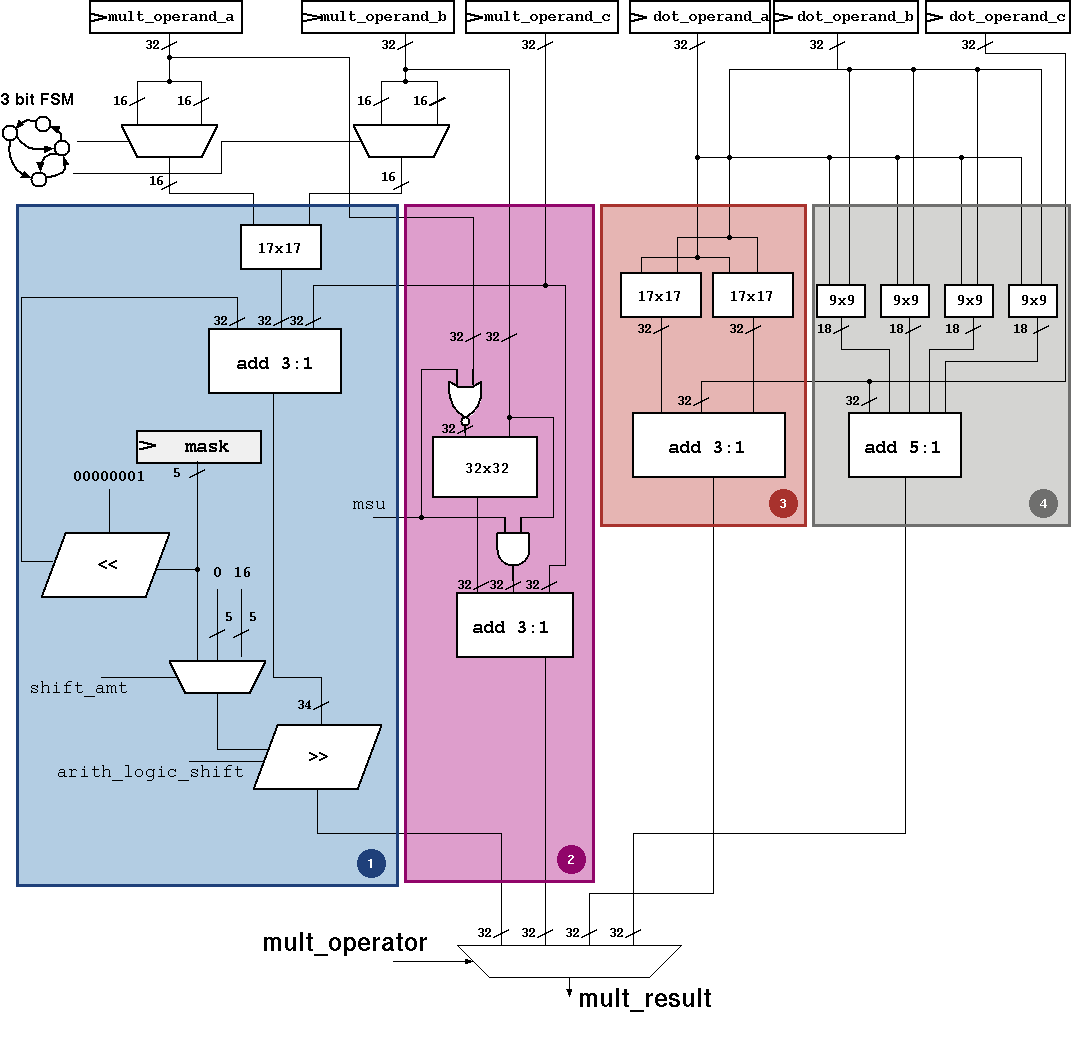}
    \caption{Simplified block diagram of the Multiplier in the \riscv core implementation.}
    \label{fig:mult_archi}
\end{figure}

While adding vectorial support for add/sub and logic operations was achieved with relative ease, the design of the multiplier was more involved. The final multiplier shown in \figref{fig:mult_archi} contains three modules:  A 32b$\times$32b multiplier, a fractional multiplier, and two dot-product (dotp) multipliers.

The proposed multiplier, has the capability to multiply two vectors and accumulate the result in a 32b value in one cycle. A vector can contain two 16b elements or four 8b elements. To perform signed and unsigned multiplications, the 8b/16b inputs are sign extended, therefore each element is a 17b or 9b signed word. A common problem with an $N$ bit multiplier is that its output needs to be $2\cdot N$ bits wide to be able to cover the entire range. In some architectures an additional register is used to store part of the multiplication result. The dot product operation produces a result with a larger dynamic than its operands without any extra register due to the fact that its operands are either 8b or 16b.
Such dot-product\,(\emph{dotp}) operations can be implemented in hardware with four multipliers and a compression tree and allow to perform up to four multiplications and three additions in a single operation as follows:
\begin{equation*}
d = a[0]\cdot b[0] + a[1]\cdot b[1] + a[2]\cdot b[2] + a[3]\cdot b[3] ,
\end{equation*}
where a[i], b[i] are the individual bytes of a register and d is the 32b accumulation result. The multiply-accumulate\,(MAC) equivalent is the Sum-of-Dot-Product\,(\emph{sdotp}) operation which can be implemented with an additional accumulation input at the compression tree.
With a vectorized ALU, and \emph{dotp}-operations it is possible to significantly increase the computational throughput of a single core when operating on short binary numbers.

The implementation of the dot-product unit has been designed such that its longest path is shorter or equal to the critical path of the overall system. In our case, this path is from the processor core to the memories and vice versa. This has led to a design where the additional circuitry to support the vector operations did not have an impact on the overall operation speed. The dot-product-units within the multiplier are essentially split in separate  16b dotp-unit and a 8b dotp-unit. \figref{fig:mult_archi} shows the 16b dotp-unit (region \textit{3}) and 8b dotp-unit (region \textit{4}) which have been implemented by one partial product compressor which sums up the accumulation register and all partial products coming from the partial product generators. The multiplications exploit carry-save format without performing a carry-propagation before the additions.

The proposed multiplier also offers functionality to support fixed-point numbers. In this mode the \emph{p.mul} instruction accepts two 16b values (signed or unsigned) as input operands and calculates a 32b result. The \emph{p.mac} multiply-add instruction allows an additional 32b value to be accumulated to the result. Both instructions produce a 32b value which can be shifted to the right by $I$ bits, moreover it is possible to round the result (adding $2^{I-1}$) before shifting as shown in \figref{fig:mult_archi} where the fractional multiplier is shown in region \textit{1}. 

Consider the code example given in \tabref{tab:mulQ211} that demonstrates the difference between the \riscv ISA with and without fixed-point multiplication support. In this example, two vectors of \textit{n} elements containing Q1.11 elements are multiplied with each other (a common operation in the frequency domain to perform convolution). The multiplication will result in a number expressed in the format Q2.22 and a subsequent rounding and normalization step will be needed to express the result using 12 bits as a Q2.10 number. It is important to note that, for such operations, performing the rounding operation before normalization reduces the error.

\begin{table}[tb]
\small
\begin{threeparttable}
\renewcommand\arraystretch{0.3}
\caption{Element-wise multiplication of \textit{n} Q1.11 elements with round and normalization.}
\centering
\label{tab:mulQ211}
\begin{tabular*}{0.49\textwidth}{l@{\extracolsep{\fill}}*{2}{l}}
\toprule
Without Mul Norm Round & With Mul Norm Round \\
\midrule
  \texttt{addi ~~~~~r3, r0, \textit{n} } & \texttt{addi ~~~~~r3, r0, \textit{n} }  \\
  \texttt{lp.setupi r0, r3, endL  } & \texttt{lp.setupi r0, r3, endL  }  \\
  \texttt{p.lh ~~~~~0(r10!), r4        } & \texttt{p.lh ~~~~~0(r10!), r4        }  \\
  \texttt{p.lh ~~~~~0(r11!), r5        } & \texttt{p.lh ~~~~~0(r11!), r5        }  \\
  \texttt{mul  ~~~~~~r4, r4, r5         } & \texttt{p.mulsRN ~r4, r4, r5, 12 }  \\
  \texttt{addi ~~~~~r4, r4, 0x800      } & \texttt{endL: sw ~0(r12!), r4    }  \\
  \texttt{srai ~~~~~r4, r4, 12         } & \texttt{                        }  \\
  \texttt{endL: sw ~0(r12!), r4    } & \texttt{                        }  \\
\bottomrule
\end{tabular*}
\end{threeparttable}
\end{table}

The \emph{p.mulsRN} multiply-signed with round and normalize operation is able to perform all three operations (multiply, add and shift) in one cycle thus reducing both the codesize and the number of cycles. The two additional 32b values need to be added to the  partial-products compressor, which does not increase the number of the levels of the compressor-tree used in in the multiplier and therefore does not play a major factor in the overall delay of the circuit\,\cite{Parhami1996}.

Naturally the multiplier also supports standard 32b$\times$32b integer multiplications, and it is possible to also support 16b$\times$16b + 32b multiply-accumulate operations at no additional cost. Similar to the ARM Cortex M4 \textit{MLS} instruction a multiply with subtract using 32b operands is also supported.

The proposed ISA-extensions are realized with separate execution units in the EX-stage which have contributed to an increase in area (8.3\,kGE ALU, 12.6\,kGE multiplier). To keep the power consumption at a minimum, switching activity at unused parts of the ALU has to be kept at a minimum. Therefore, all separate units: the ALU, the integer and fractional multiplier, and the dot-product unit all have separate input operand registers which can be clock gated. The input operands are controlled by the instruction decoder and can be held at constant values to further eliminate propagation of switching activity in idle units.
The switching activity reduction is achieved by additional flip-flops\,(FF) at the input of each unit (192 in total). These additional FFs allow to reduce the switching activity which decreases the power consumption of the core by 50\%.

%% file: tabisa.tex
\begin{table*}[tb]
\small
\begin{threeparttable}
\renewcommand\arraystretch{0.3}
\caption{Instructions Description}
\centering
\label{tab:newinstructions}
\begin{tabularx}{\textwidth}{p{4.4cm}p{3.7cm}|p{4.4cm}p{3.7cm}}
\toprule
      \centerline{\textbf{Instruction format}} & \centerline{\textbf{Description}}         & \centerline{\textbf{Instruction format}} & \centerline{\textbf{Description}} \\
      \toprule
      \multicolumn{2}{c|}{\textbf{Hardware Loop Instructions}}                             & \multicolumn{2}{c}{\textbf{Fixed Point Instructions}}  \\
      \midrule
      \texttt{lp.starti ~L, ~I } & Set the HW loop start address  & \texttt{p.add\lbrack R\rbrack N ~~~~rD, rA, rB, I} & Addition with round and normalization by I bits $^a$ \\
      \texttt{lp.endi ~~~L, ~I } & Set the HW loop end address    & \texttt{p.sub\lbrack R\rbrack N ~~~~rD, rA, rB, I} & Subtraction with round and normalization by I bits $^a$\\
      \texttt{lp.count ~~L, ~rA} & Set the HW loop number of iterations & \texttt{p.mul\lbrack hh\rbrack\lbrack R\rbrack N rD, rA, rB, I} & Multiplication with round and normalization by I bits $^a$$^b$\\
      \texttt{lp.setup ~~L, ~rA, ~~I} & HW loop setup with registers  & \texttt{p.mac\lbrack hh\rbrack\lbrack R\rbrack N rD, rA, rB, I} & MAC with round and normalization by I bits $^a$$^b$ \\
      \texttt{lp.setupi ~L, ~I1, ~~I2} & HW loop setup with immediate   & \texttt{p.clip ~~~~~~~rD, rA, I} & Clip the value between $-2^{I-1}$ and $2^{I-1}-1$ \\
      \toprule
      \multicolumn{2}{c}{\textbf{Extended Load/Store Instructions}}   & \multicolumn{2}{c}{\textbf{Vectorial Instructions}}\\
      \midrule
      \texttt{p.l\{b,h,w\} rD, \{rB,I\}(rA)} & Load a value from address (rA+\{rB,I\})$^c$  & \texttt{pv.inst.\{b,h\} rD, rA, rB} & General vectorial instruction between two registers $^c$\\
      \texttt{p.l\{b,h,w\} rD, \{rB,I\}(rA!)}    & Load a value from address rA and increment rA by \{rB,I\} $^c$ & \texttt{pv.inst.\{b,h\} rD, rA, I} & General vectorial instr. between a register and an immediate $^c$\\
      \texttt{p.s\{b,h,w\} rB, \{rD,I\}(rA)}     & Store a value to address (rA+\{rD,I\})              $^c$ & & \\
      \texttt{p.s\{b,h,w\} rB, \{rD,I\}(rA!)}    & Store a value to address rA and increment rA by \{rD,I\} $^c$ & & \\
      \bottomrule
\end{tabularx}
\begin{tablenotes}
\item[$^a$] If R is not specified, there is no round operation before shifting.
\item[$^b$] If hh is specified, the operation takes the higher 16b of the operands.
\item[$^c$] b, h, w specific the data lenght of the operands: byte (8b), halfword (16b), word (32b).
\end{tablenotes}
\end{threeparttable}
\end{table*}

%% file: 05_sw.tex
The PULP-compiler used in this work has been derived from the original GCC \riscv version which is itself derived from the MIPS GCC version. GCC-5.2 release and the latest binutils release have been used.
The binutils have been modified to support the extended ISA as well as a few new relocation schemes have been added.

Hardware loop detection and mapping has been enabled as well as post modified pointers detection. The GCC internal support for hardware loops is sufficient but the module taking care of post modified pointers is rather old and primitive. As a comparison, more recent modules geared toward vectorization and dependency analysis are way more sophisticated. One of the consequences is that the scope of induced pointers is limited to a single loop level and we miss opportunities that can be exposed across loop levels in a loop nest.

Further, sum of products and sum of differences are automatically exposed allowing the compiler to always favor the shortest form of mac/msu (use 16b$\times$16b into 32b instead of 32b$\times$32b into 32b) to reduce energy.

Vector support (4 bytes or 2 shorts) has been enabled to take advantage of the SIMD extensions. The fact that we support unaligned memory accesses plays a key role into the exposition of vectorization candidates.
Even though auto-vectorization works well, we believe that for deeply embedded targets such as PULP, it makes sense to use the GCC extensions to manipulate vectors as C/C++ objects. It avoids overly conservative prologue and epilogue insertion created by the auto vectorizer that are having a serious negative impact on code size. We have opted for a less specialized scheme for the fixed-point representation, to give more flexibility. The architecture supports any fixed-point format between Q2 and Q31 with optimized instructions for normalization, rounding and clipping.

All these instructions fit nicely into the instruction combiner pass of GCC. Normalization and rounding can be combined with common arithmetic instructions such as addition, subtraction and multiplication/mac, everything being performed in one cycle.
Dot product instructions are more problematic since they are not a native GCC internal operation and the depth of its representation prevent it from being automatically detected and mapped by the compiler which is the reason why we rely on built-in support.
More generally most of our extensions can also be used through built-ins as an alternative to automatic detection (in contrast with assembly insertions built-ins can easily capture the precise semantic of the instructions they implement). An example of using \emph{dotp} instructions is given below:

\definecolor{mygreen}{rgb}{0,0.6,0}
\definecolor{mygray}{rgb}{0.5,0.5,0.5}
\definecolor{mymauve}{rgb}{0.58,0,0.82}

\lstdefinestyle{customc}{
  belowcaptionskip=2\baselineskip,
  breaklines=true,
  frame=single,
  language=C,
  showstringspaces=false,
  basicstyle=\footnotesize\ttfamily,
  keywordstyle=\bfseries\color{green!40!black},
  commentstyle=\itshape\color{purple!40!black},
  basicstyle=\footnotesize\ttfamily,
  stringstyle=\color{orange},
}
\lstset{escapechar=@,style=customc}

\begin{lstlisting}
 // define vector data type and dotp instruction
 typedef short PixV __attribute__((vector_size(4)));
 #define SumDotp16(a,b,c) __builtin_sdotsp2(a,b,c);

 PixV VectA, VectB; // vectors of shorts
 int S;
 ...
 S = 0;
 // each iteration is computing two mult and 2 accum
 for (int k = 0; k < (SIZE>>1); k++) {
   S = SumDotp16(VectA[k], VectB[k], S);
 }
 C[i*N+j] = S;
 ...
\end{lstlisting}

Finally, the bit manipulation part of the presented ISA-extensions fits well into GCC since most instructions have already an internal GCC counterpart.
\footnote{ PULP Parallel Programming is possible thanks to openMP 3 support integrated in our GCC compiler. The interested reader is referred to ~\cite{Rossi2014b} for more information.}

%% file: 06_results.tex
For hardware, power and energy efficiency evaluations, we have implemented the original and extended core in a PULP-cluster with 72\,kB TCDM memory and 4\,kB I\$. The two clusters\,(cluster A with a RVC32IM \riscv core, and cluster B with the same core plus the proposed extensions) have been synthesized with Synopsys Design Compiler-2016.03 and complete back-end flows have been performed using Cadence Innovus-15.20.100 in a 8-metal UMC 65\,nm LL CMOS technology. A set of benchmarks has been written in C (with no assembly-level optimization) and compiled with the modified \riscv GCC toolchain which makes use of the ISA-extensions.
The switching activity of both clusters have been recorded from simulations in Mentor QuestaSim 10.5a using back-annotated post-layout gate-level netlists and analyzed with the obtained value change dump\,(VCD) files in Cadence Innovus-15.20.100.

In the following sections, we will first discuss the area, frequency and power of the cluster in \secref{sec:areapower}. Then the energy efficiency of several instructions is discussed in \secref{sec:instruction}. Speedup and energy efficiency gains of cluster B are presented in \secref{sec:benchmarks} followed by an in-depth discussion about convolutions in \secref{sec:conv_perf} including a comparison with state-of-the-art hardware accelerators. While the above sections focus on relative energy, and power comparisons which are best done in a consolidated technology in super threshold conditions, the overall cluster performance is analyzed in NT conditions and compared to previous PULP-architectures in \secref{sec:pulpcomp}.

\subsection{Area, Frequency, and Power}
\label{sec:areapower}

\begin{figure}[tb]
\centering
  \includegraphics[width=0.47\linewidth]{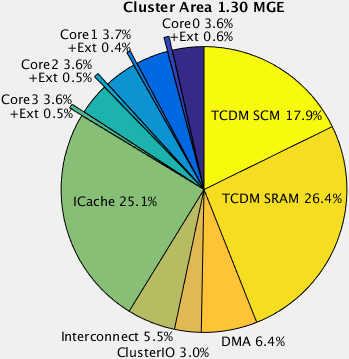}
  \includegraphics[width=0.50\linewidth]{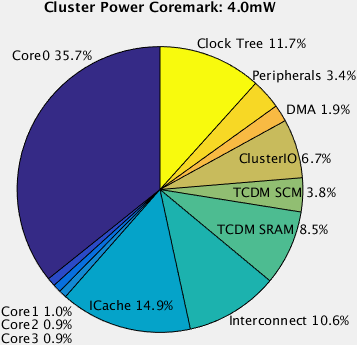}
  \caption{a) Area distribution of the cluster optimized for 2.8\,ns cycle time. The DSP-extensions of each core are highlighted. b) Power distribution when running CoreMark on a single core at 50\,MHz, 1.08\,V.}
  \label{fig:clusterpie}
\end{figure}

\input{\tabpath/armcomp.tex}

\tabref{tab:armcomp} shows a comparison of the basic and extended \riscv cores with an OpenRISC architecture, and the ARM Cortex M4. It can be seen that the \riscv is very similar in performance to OpenRISC architecture, but with a smaller dynamic power consumption due to support of compressed instructions\,(less instruction fetches) and low-level power optimizations to reduce switching activity.
The extended \riscv core architecture increases in size by 6.6\,kGE as a result of additional execution units (dot-product unit, shuffle unit, fixed-point support) but these extensions allows it to reach a significantly higher CoreMark score of 2.84. 

With respect to ARM architectures, the proposed core is comparable in size, but can not reach the CoreMark score due to branch costs, cache misses and load-hazards. This is mainly due to the fact that CoreMark contains significant amount of control-intensive code which cannot be accelerated by SIMD-extensions. The real performance of the proposed core becomes apparent in a shared-memory multi-core system, where even in a four-core configuration it reaches higher operating frequencies on comparable technologies. 

The area distribution of the cluster is shown in \figref{fig:clusterpie}a). The total area of the cluster is 1.30\,MGE of which the additional area for the dot-product unit and ALU-extensions accounts for only 2\%. \figref{fig:clusterpie}b) shows the power distribution when running CoreMark on a single core. The total power consumption at 50\,MHz and 1.08V is only 4\,mW of which 35\% is consumed by the active core. The three idle cores consume only 2.8\% of which 83\% is leakage.

\subsection{Instruction Level Performance}
\label{sec:instruction}

\begin{figure}[b!]
\centering
  \includegraphics[width=1\linewidth]{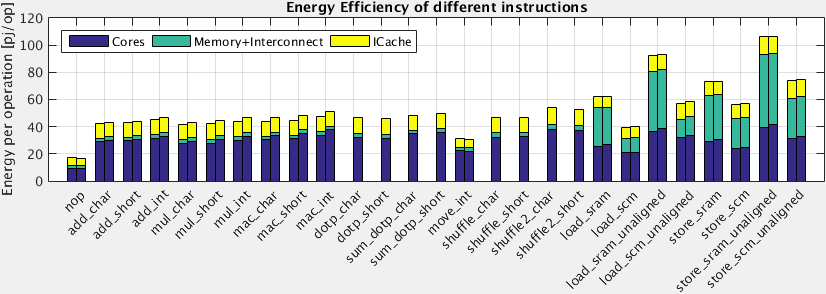}
  \caption{Energy per operation for different instructions with random inputs, 1.08V, worst case conditions, 65\,nm CMOS.}
  \label{fig:insn_level_perf}
\end{figure}

The power consumption and energy efficiency of the two \riscv core versions have been analyzed at instruction level. To determine the power and energy of each instruction, they have been executed in a loop with 100 iterations each containing 100 instructions. The energy of a single instruction is then the product of the execution time and the power divided by the number of executed instructions (10'000) and cores (4). The power of each instruction consists of TCDM-power (including interconnect), core-power, and I\$-power. The execution time of the loop depends on the latency of the instruction (2 cycles for unaligned memory access instructions, 1 cycle for others).

\begin{figure*}[tb]
\centering
  \includegraphics[width=0.48\linewidth]{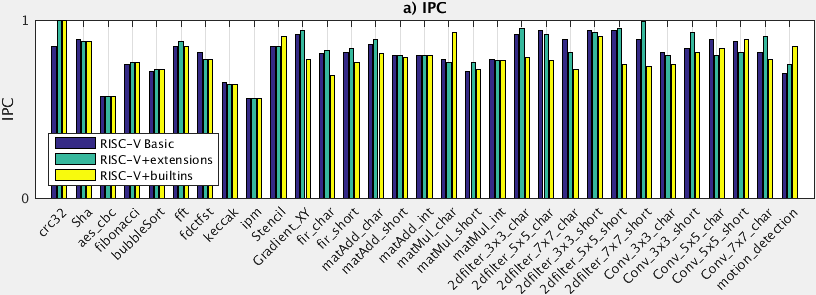}
  \includegraphics[width=0.48\linewidth]{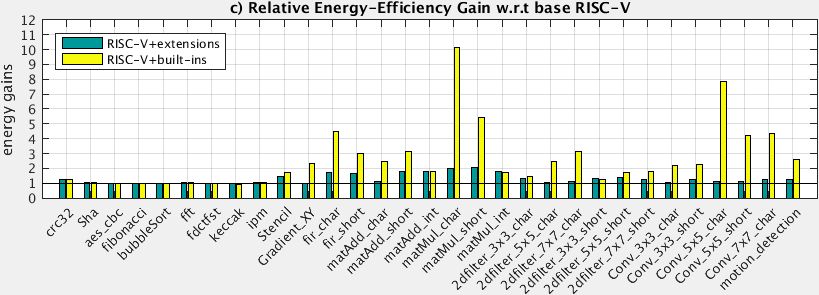}
  \includegraphics[width=0.48\linewidth]{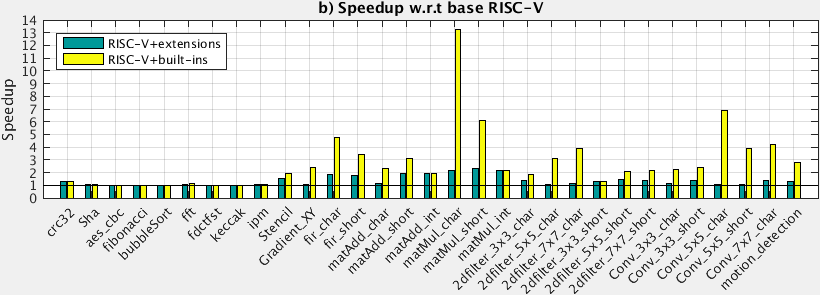}
  \includegraphics[width=0.48\linewidth]{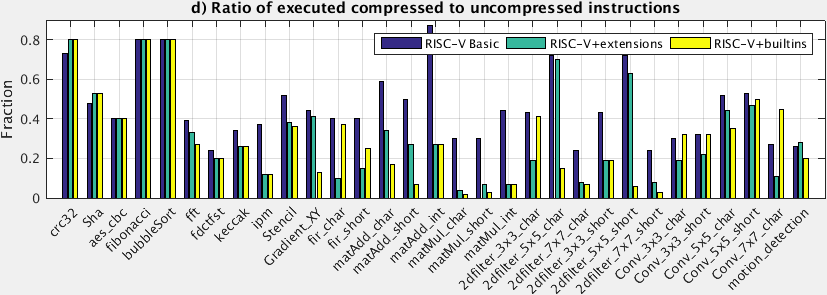}
  \caption{a) IPC of all benchmarks with and without extensions, and with built-ins, b) Speedup, c) energy efficiency gains with respect to a \riscv cluster with basic extensions, d) Ratio of executed instructions (compressed/normal).}
  \label{fig:bench_results}
\end{figure*}

\figref{fig:insn_level_perf} shows the resulting energy per operation numbers for different types of instructions. As expected,  \emph{nop} consumes the least power, which is not 0, because it has to be fetched from the I\$/L0-buffer. Then comes a set of arithmetic instructions (\emph{add, mul, mac, dotp, sdotp}) for different data types (char, short, int). The extended core is bigger and hence, it also consumes more power. For the arithmetic operations we observe a slight power increase of 4\%.

While some of the new instructions (\emph{mac}, \emph{sdotp}, \emph{shuffle})consume slightly more power than other arithmetic instructions, they actually replace several instructions and are actually more energy efficient. For example, the \emph{shuffle} instruction is capable of reorganizing bytes or shorts in any combination by utilizing the shuffle-unit. Executing a shuffle operation (50\,pJ) is equivalent to executing 3-4 simple ALU operations for 90-120\,pJ. Similarly the proposed LSU allows to perform unaligned memory accesses from SRAMs for only 93-106\,pJ whereas software only support would require to execute 5 instructions in sequence which would cost about 3$\times$ as much energy. As a final observation we note that \emph{ld/st} operations from SCMs are on average 46\% more energy-efficient than from SRAMs. Unfortunately, SCMs are not very area efficient for larger memories and are therefore limited in size. 

\begin{figure*}[tb]
\centering
  \includegraphics[width=0.32\linewidth]{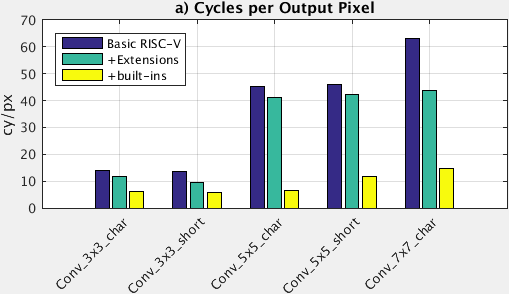}
  \includegraphics[width=0.32\linewidth]{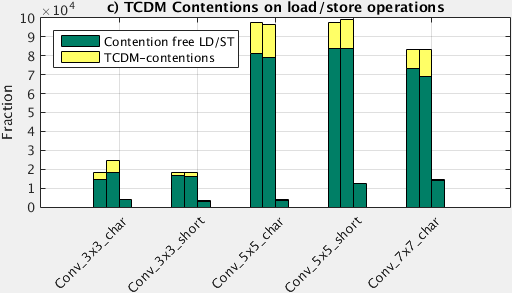}
  \includegraphics[width=0.32\linewidth]{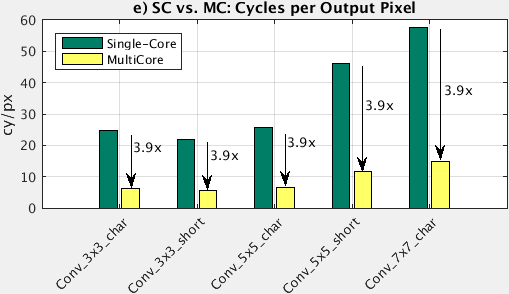}
  \includegraphics[width=0.32\linewidth]{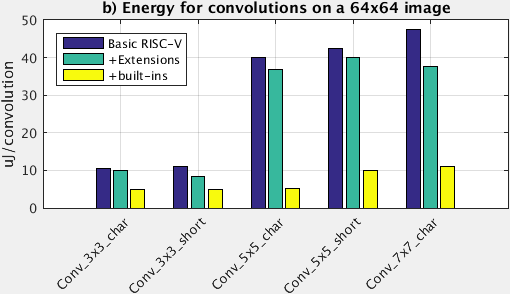}
  \includegraphics[width=0.32\linewidth]{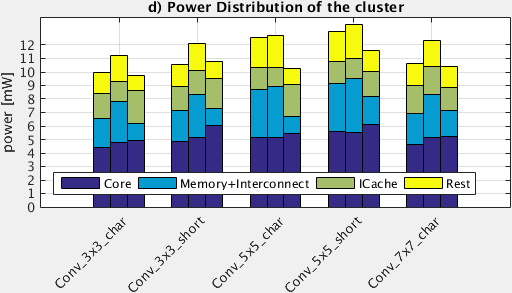}
  \includegraphics[width=0.32\linewidth]{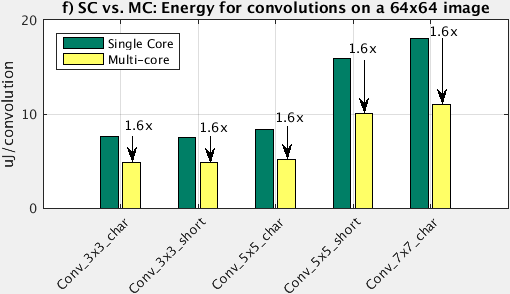}
  \caption{a) required cycles per output pixel, b) total energy consumption when processing the convolutions at 50\,MHz at 1.08V, c) number of TCDM-contentions,  and d) the power distribution of the cluster when using the three diffrent instruction sets. e) and f) compare single- and multi-core implementations in speedup and energy.}
  \label{fig:conv_res}
\end{figure*}

\subsection{Function Kernel Performance}
\label{sec:benchmarks}

To evaluate the performance gain of the proposed extensions, a set of benchmarks ranging from cryptographic kernels\,(crc, sha, aes, keccak), control intensive applications\,(fibonacci, bubblesort), transformations (FFT, FDCT), over more data intensive linear algebra kernels\,(matrix additions, multiplications), to various filters\,(fir, 2D filters, convolutions) has been compiled. We have also evaluated a \emph{motion detection} application which makes use of linear algebra, convolutions and filters.

\figref{fig:bench_results}a) shows the IPC of all applications and \figref{fig:bench_results}b) shows the speedup of the \riscv core with hardware loops and post-increment extensions, versus a plain \riscv ISA. In this case an average speedup of 37\% can be observed. As expected, filters and linear algebra kernels with very regular data access patterns benefit the most from the extensions.
A second bar\,(+built-ins), shows the full power of the proposed \riscv ISA-extensions. Data intensive kernels benefit from vector extensions, which can be used with vector data types and the C built-in \emph{dotp}-instruction. On these kernels a  much larger speedup, up to 13.2$\times$, can be achieved, with an average gain of 3.5$\times$.

The overall energy gains of the extended core are shown in \figref{fig:bench_results}c) which shows an average of 3.2$\times$ gains. In an ideal case, when processing a matrix multiplication that benefits from \emph{dotp}, hardware loop and post-increment extensions the gain performance gain can reach 10.2$\times$.

Finally, the ratio of executed instructions (compressed/normal) is shown in \figref{fig:bench_results}d). We see that 28.9-46.1\% of all executed instructions were compressed. Note that, the ratio of compressed instructions is less in the extended core, as vector instructions, and built-ins do not exist in compressed form.

\subsection{Convolution Performance}
\label{sec:conv_perf}

\input{\tabpath/ConvComp}

Convolutions are common operations and are widely used in image processing. In this section we compare the performance of \emph{basic} \riscv implementation to the architecture with the \emph{extensions} proposed in this paper. Since not all instructions can be efficiently utilized by the compiler, we also provide a third comparison called \emph{built-ins} which runs on the extended architecture by directly calling these extended instructions. For the convolutions, a 64$\times$64 pixel image has been used  with a Gaussian filter of size 3$\times$3, 5$\times$5, and 7$\times$7. Convolutions are shown for 8b (char) and 16b (short) coefficients.

\figref{fig:conv_res}a) shows the required cycles per output pixel using the three different versions of the architecture. In this evaluation, the image to be processed is divided into four strips and each core working on its own strip. Enabling the extensions of the core allows to speedup the convolutions by up to 41\% mainly due to the use of hardware loops, and post-increment instructions. Another 1.7-6.4$\times$ gain can be obtained when using the \emph{dotp} and \emph{shuffle} instructions resulting in an overall speedup of 2.2-6.9$\times$ on average. \emph{Dotp} instructions are processing four multiplications, three additions and an accumulation all within a single cycle and can therefore reduce the number of arithmetic instructions significantly. A 5$\times$5 filter would require 25 \emph{mac} instructions, while 7 \emph{sdotp} instructions are sufficient when the vector extensions are used. As seen in
\figref{fig:conv_res}b) the acceleration also directly translates into energy savings, the extended architecture computing convolutions on a 64$\times$64 image is 2.2-7.8$\times$ more energy efficient.

Utilizing vector extensions and the RF as a L0-storage allows to reduce the number of load words as shown in \figref{fig:conv_res}c) because it is possible to keep more coefficients in the RF. When computing 3$\times$3 and 5$\times$5 convolutions it is even possible to store all required elements in the RF, and only load the new pixels. The remaining pixels can be reorganized using \emph{move} and \emph{shuffle} instructions as described in the 2D convolution example of \secref{sec:pulp}. This not only reduces the number of \emph{ld/st} operations by 8.3$\times$, but also reduces contentions at the 8-bank shared memory. The extended cores have a higher \emph{ld/st}-density and thus experience the more contentions per TCDM access, 17.8\% on average. With vector operations this number goes down to only 6.2\% which is a reduction from 11'100 contentions to only 390.

\figref{fig:conv_res}d) shows the power distribution of the cluster. It is interesting to note that, although for all examples the power of the core increases, the overall system power is reduced in all but one case, where it marginally increases by 5.3\% (\emph{Conv\_3x3\_short}).

The speedup and energy saving of the vector convolutions on the four core system are shown in \figref{fig:conv_res}e) and f) with respect to a single-core configuration. Overheads at the strip boundaries of the multi-core implementation are negligible as the speedup with 3.9$\times$ is almost ideal. Using four instead of one core requires only 2.4$\times$ more power leading to energy savings of 1.6$\times$. Hence, the system is very well scalable both in execution speed and energy.

\tabref{tab:hwacc} shows a comparison of the proposed \riscv core with its plain ISA and state-of-the art hardware accelerators. Origami\,\cite{Cavigelli2016} is a convolutional network accelerator capable of computing 7$\times$7 convolutions and generating 4 output pixels per cycle. Its area with 912\,kGE is almost the size of a complete cluster. HWCE\,\cite{Conti2015} is an accelerator which can be plugged to a memory system and is approximately 184\,kGE big and can compute almost 2 pixels per cycle. Typically, a hardware accelerator outperforms a general purpose architecture by factors of 100-1000. This certainly holds true for the \riscv core without extensions, which is  112$\times$ slower than the HWCE, while the proposed \riscv core with DSP extensions is only 11-26$\times$ slower. This is a significant step in closing the gap between programmable architectures and specialized hardware.

 In terms of energy the \riscv core consumes only 15-25$\times$ more than the Origami chip which is implemented in the same technology. 

\begin{figure}[tb]
\centering
  \includegraphics[width=1.00\linewidth]{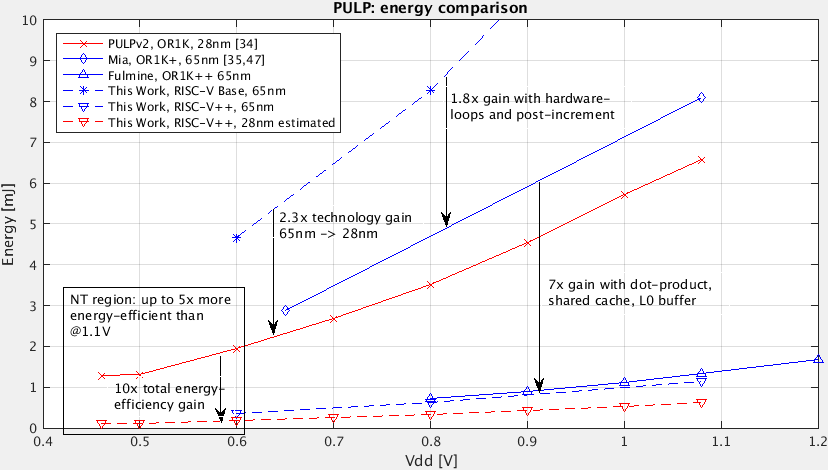}
  \caption{Energy consumption of different PULP architectures when processing a matrix multiplication on four cores that shows the potential of the vector externsions.}
  \label{fig:pulpcomp}
\end{figure}

\subsection{Near-threshold operation}
\label{sec:pulpcomp}

The ISA extensions of the \riscv cores decrease the execution time significantly which at the same time reduces the energy. \figref{fig:pulpcomp} shows the energy efficiency of several successfully taped-out PULP chips when processing a matrix multiplication on four cores. First, PULPv2\,\cite{Rossi2016a} an implementation of PULP with basic OpenRISC cores without ISA-extensions in 28\,nm FDSOI and second Mia\,\cite{Pullini2016} a PULP cluster in 65\,nm with an OpenRISC core featuring a first set of ISA-extensions (hardware-loops, post-increment instructions)\,\cite{Gautschi2015}. While Mia had no support for dot-products, and shuffle, Fulmine is already supporting a very similar ISA as the presented RISC-V core in this work. All chips are silicon proven and measured.

PULPv2, Mia, and this work feature SCMs which allow for NT-operation. PULPv2 for example works down to 0.46V where it consumes only 1.3\,mJ for a matrix multiplication which is 5$\times$ less than at 1.1V.

The plot in \figref{fig:pulpcomp} shows the evolution of the architecture. First we observe a 2.3$\times$ gain the 28\,nm PULPv2 versus a 65\,nm RISC-V based cluster with comparable instruction sets (both architectures feature no ISA-extensions). Then we see that the introduction of hardware loops, and post-increment instructions in Mia almost closes this gap as the execution speed improved by 1.8$\times$.
The ISA-extensions presented in this paper bring the largest speedup due to the introduction of dot-product instructions and accounts for a factor of 10$\times$ with respect to a \riscv or OpenRISC core without extensions.

The graph also shows the expected energy efficiency of these extensions in an advanced technology like 28\,nm FDSOI. These estimations show that moving to an advanced technology and operating at the NT region will provide another 1.9$\times$ advantage in energy consumption with respect to the 65\,nm implementation reported in this paper.
A 28\,nm implementation of the RISC-V cluster will work in the same conditions (0.46\,V-1.1\,V), achieve the same throughput (0.2-2.5\,GOPS) while consuming 10$\times$ less energy due to the reduced runtime and only moderate increase in power. The increased efficiency, as well as low power consumption (1\,mW at 0.46\,V) and still high computation power (0.2\,GOPS at 0.46\,V, 40\,MHz) allows the cluster to be perfectly suited for data processing in IoT endpoint devices.

%
%
%
%
%
%
%
%
%

%% file: armcomp.tex
\begin{table}[t]
\begin{threeparttable}
\renewcommand{\arraystretch}{0.9}
\caption{Comparison of different core architectures}
\label{tab:armcomp}
\centering
\begin{tabularx}{\columnwidth}{@{}l|p{1.0cm}|p{1.0cm}|p{1.2cm}|p{0.8cm}p{0.8cm}@{}}
\toprule
 & \multicolumn{2}{c|}{\textbf{This Work}}                        &  & \multicolumn{2}{c}{}\\
\textbf{Processor Core} & \textbf{RISC-V Basic} & \textbf{RISC-V Ext.}    & \textbf{OpenRISC} \cite{Gautschi2015}   & \multicolumn{2}{c@{}}{\textbf{CortexM4} \cite{CortexM4}}  \\
\midrule                                                                                                                                                          
Technology                     & 65\,nm      & 65\,nm      & 65\,nm      & 90\,nm       & 40\,nm         \\
\midrule
Vdd\,[V]                       & 1.08        & 1.08        & 1.2         & 1.2          & 1.1            \\
Freq.\,[MHz]                   & 357         & 357         & 362         & 216          & 216            \\
Dyn. Power                    & \multirow{2}*{26.28}       & \multirow{2}*{28.68}       & \multirow{2}*{33.8}        & \multirow{2}*{32.8}         & \multirow{2}*{12.26}          \\
\,[\textmu W/MHz]              &             &             &             &              &              \\
Area\,[kGE]                    & 46.9        & 53.5        & 44.5        & -            & -              \\
\,[mm$^2$]                 & 0.068       & 0.077       & 0.064       & 0.119        & 0.028          \\
\midrule
CoreMark/MHz                   & 2.43        & 2.84        & 2.66        & \multicolumn{2}{c}{3.40}      \\

\bottomrule
\end{tabularx}
\end{threeparttable}
\end{table}

%% file: ConvComp.tex
\begin{table*}[tb]
\begin{threeparttable}
\renewcommand*{\arraystretch}{0.8}
\caption{Comparison of convolutional performance with hardware accelerators}
\label{tab:hwacc}
\centering
\begin{tabularx}{\textwidth}{@{}lp{1.5cm}p{1.5cm}p{1.5cm}p{1.5cm}p{1.5cm}p{1.5cm}p{1.5cm}p{1.5cm}@{}}
\toprule
 & \multicolumn{4}{c}{\textbf{This Work}} & \multicolumn{2}{c}{\textbf{Origami~\cite{Cavigelli2016}}} & \multicolumn{2}{c@{}}{\textbf{HWCE~\cite{Conti2015}}}\\
\textbf{Type}              & \multicolumn{2}{@{}c}{RISC-V Base ISA}& \multicolumn{2}{@{}c}{RISC-V + DSP-Ext.} & \multicolumn{2}{@{}c}{Accelerator} & \multicolumn{2}{@{}c}{Accelerator} \\
\midrule
\textbf{Implementation}    & & & & & & & & \\
Technology                 & \multicolumn{2}{@{}c}{UMC 65nm}      & \multicolumn{2}{@{}c}{UMC 65nm}       & \multicolumn{2}{@{}c}{UMC 65nm}   & \multicolumn{2}{@{}c}{ST FDSOI 28nm} \\
Results from:              & \multicolumn{2}{@{}c}{Post-layout}   & \multicolumn{2}{@{}c}{Post-layout}    & \multicolumn{2}{@{}c}{Silicon}    & \multicolumn{2}{c}{Post-layout}      \\
Coef. size                 &  \multicolumn{2}{@{}c}{8b/16b/32b}   &  \multicolumn{2}{@{}c}{8b/16b/32b}    &  \multicolumn{2}{@{}c}{12b}       &  \multicolumn{2}{@{}c}{16b}          \\
Core Area [kGE]            & \multicolumn{2}{@{}c}{$4\times 46.9$}& \multicolumn{2}{@{}c}{$4\times 53.5$ }& \multicolumn{2}{@{}c}{- }         & \multicolumn{2}{@{}c}{184$^b$}       \\
Total Area [kGE]           & \multicolumn{2}{@{}c}{1277}          & \multicolumn{2}{@{}c}{1300 }          & \multicolumn{2}{@{}c}{912 }       & \multicolumn{2}{@{}c}{-}             \\
Supply Voltage [V]         & \multicolumn{1}{@{}c}{1.08V} & \multicolumn{1}{@{}c}{0.6V}      &  \multicolumn{1}{@{}c}{1.08V} & \multicolumn{1}{@{}c}{0.6V}                       & \multicolumn{1}{@{}c}{1.2V}   & \multicolumn{1}{@{}c}{0.8V}     & \multicolumn{1}{@{}c}{0.8V}    & \multicolumn{1}{@{}c}{0.4V}                       \\
Frequency [MHz]            & \multicolumn{1}{@{}c}{357}   & \multicolumn{1}{@{}c}{50$^c$}                       &  \multicolumn{1}{@{}c}{357}   & \multicolumn{1}{@{}c}{50$^c$}                   & \multicolumn{1}{@{}c}{500}     &  \multicolumn{1}{@{}c}{189}                     & \multicolumn{1}{@{}c}{400}      & \multicolumn{1}{@{}c}{22}       \\
\midrule
\textbf{3x3 Convolution}            & & & & & & & & \\
Performance [Cycles/px]    &  \multicolumn{1}{@{}c}{14.0}     & \multicolumn{1}{@{}c}{14.0}     & \multicolumn{1}{@{}c}{6.3}      &\multicolumn{1}{@{}c}{6.3}      & \multicolumn{1}{@{}c}{-}       &  \multicolumn{1}{@{}c}{-}                   & \multicolumn{1}{@{}c}{0.56}     & \multicolumn{1}{@{}c}{0.56}     \\
Energy efficiency [pJ/px]  &  \multicolumn{1}{@{}c}{2570}     & \multicolumn{1}{@{}c}{839$^c$}  & \multicolumn{1}{@{}c}{1179}     & \multicolumn{1}{@{}c}{384$^c$}   & \multicolumn{1}{@{}c}{-}       & \multicolumn{1}{@{}c}{-}                    & \multicolumn{1}{@{}c}{-}        &  \multicolumn{1}{@{}c}{20$^a$}   \\
\midrule
\textbf{5x5 Convolution}            & & & & & & & & \\
Performance [Cycles/px]    & \multicolumn{1}{@{}c}{45.1}      & \multicolumn{1}{@{}c}{45.1}     & \multicolumn{1}{@{}c}{6.6}      & \multicolumn{1}{@{}c}{6.6}      & \multicolumn{1}{@{}c}{-}       &  \multicolumn{1}{@{}c}{-}                   & \multicolumn{1}{@{}c}{0.56}     & \multicolumn{1}{@{}c}{0.56}     \\
Energy efficiency [pJ/px]  & \multicolumn{1}{@{}c}{10094}     & \multicolumn{1}{@{}c}{3261$^c$} & \multicolumn{1}{@{}c}{1286}     & \multicolumn{1}{@{}c}{418$^c$}   & \multicolumn{1}{@{}c}{-}       &  \multicolumn{1}{@{}c}{-}                   & \multicolumn{1}{@{}c}{-}        & \multicolumn{1}{@{}c}{20$^a$}   \\
\midrule
\textbf{7x7 Convolution}            & & & & & & & & \\
Performance [Cycles/px]    &  \multicolumn{1}{@{}c}{63.0}     & \multicolumn{1}{@{}c}{63.0}     & \multicolumn{1}{@{}c}{14.8}     & \multicolumn{1}{@{}c}{14.8}     & \multicolumn{1}{@{}c}{0.25}    &  \multicolumn{1}{@{}c}{0.25}                & \multicolumn{1}{@{}c}{0.56}     & \multicolumn{1}{@{}c}{0.56}     \\
Energy efficiency [pJ/px]  & \multicolumn{1}{@{}c}{12283}     & \multicolumn{1}{@{}c}{3994$^c$} & \multicolumn{1}{@{}c}{2847}     & \multicolumn{1}{@{}c}{926$^c$}   & \multicolumn{1}{@{}c}{112}     & \multicolumn{1}{@{}c}{61}                   & \multicolumn{1}{@{}c}{-}      &  \multicolumn{1}{@{}c}{20$^a$}    \\
\bottomrule
\end{tabularx}
\begin{tablenotes}
\item[$^a$] HWCE 7x7 energy, does only include accelerator.
\item[$^b$] 4.14$\times$core area (core area of 44.5\,kGE assumed\cite{Gautschi2015})
\item[$^c$] Scaled with 65nm silicon measurements.
\end{tablenotes}
\end{threeparttable}
\end{table*}

%% file: 07_conc.tex
We have presented an extended \riscv based processor core architecture which is compatible with state-of-the art cores.
The processor core has been designed for a multi-core PULP system with a shared L1 memory and a shared I\$. To increase the computational density of the processor, ISA-extensions, such as hardware loops, post-incrementing addressing modes, fixed-point and vector instructions have been added.
Powerful dot-product and sum-of-dot-products instructions on 8b and 16b data types allow to perform up to 4 multiplications and accumulations in a single cycle while consuming the same power as a single 32b MAC operation. A smart L0 buffer in the fetch stage of the cores is capable of fetching compressed instructions and buffering one cacheline, greatly reducing the pressure on the shared I\$.

On a set of benchmarks we observe that the core with the extended ISA is on average 37\% faster on general purpose applications, and by utilizing the vector extensions another gain of up to 2.3$\times$ can be achieved. When processing convolutions on the proposed core the full benefit of vector and fixed-point extensions can be used leading to an average speedup of 3.9$\times$. The use of vector instructions in combination with a L0-storage allow to decrease the shared memory bandwidth by 8.3$\times$. Since \emph{ld/st} instructions require the most energy, this decrease in bandwidth leads to energy gains up to 7.8$\times$.
The extensions allow the core to be only 15$\times$ less energy-efficient than state-of-the art hardware accelerators but are general purpose architectures and can not only be used for a specific task, but for the whole range of IoT applications.

In addition, multi-core implementations feature significantly fewer shared memory contentions with the new ISA-extensions, allowing a four-core implementation to outperform a single-core implementation by 3.9$\times$ while consuming only 2.4$\times$ more power.

Finally, implemented in an advanced 28nm technology, we observe a 5$\times$ energy efficiency gain when processing near-threshold at 0.46V where the cluster is achieving 0.2\,GOPS while consuming only 1\,mW. The cluster is scalable as it is operational from 0.46-1.1V where it consumes 1-68\,mW and achieves 0.2-2.5\,GOPS making it attractive for a wide range of IoT applications.